\pdfoutput=1

\documentclass[preprint,3p]{elsarticle}

\usepackage{lineno,hyperref}
\usepackage{bm}
\usepackage{amsmath}
\usepackage{mathtools}
\usepackage{color}
\usepackage{algorithm}
\usepackage{algorithmicx}
\usepackage{algpseudocode}

\usepackage{multirow}
\modulolinenumbers[5]

\usepackage{amssymb}

\usepackage{subfigure}
\usepackage{array}
\newcolumntype{C}[1]{>{\centering\arraybackslash}p{#1}}

\usepackage{bigints}

\usepackage{bbding}

\usepackage{units}

\usepackage{xcolor}
\usepackage{graphicx,import}

\newcommand{\Grad}[1]{\nabla {#1}}

\usepackage{stmaryrd} 

\newcommand{\avg}[1]{\{\!\{#1\}\!\}}

\newcommand{\jump}[1]{\llbracket {#1} \rrbracket }

\newcommand{\intele}[2]{ \left( {#1},{#2} \right)_{\Omega_{e}} }
\newcommand{\inteleface}[2]{ \left( {#1},{#2} \right)_{\partial\Omega_{e}} }

\newenvironment{remark}[1][Remark]{\begin{trivlist}
\item[\hskip \labelsep {\bfseries #1}]}{\end{trivlist}}

\usepackage{enumitem}
\setlist[enumerate]{label*=\roman*),ref=\roman*)}

\journal{Journal}

\begin{document}

\begin{frontmatter}

\title{Hybrid multigrid methods\\ for high-order discontinuous Galerkin discretizations}

\author[add1]{Niklas Fehn}
\ead{fehn@lnm.mw.tum.de}
\author[add1]{Peter Munch}
\ead{munch@lnm.mw.tum.de}
\author[add1]{Wolfgang A. Wall}
\ead{wall@lnm.mw.tum.de}
\author[add1]{Martin Kronbichler\corref{correspondingauthor1}}
\cortext[correspondingauthor1]{Corresponding author at: Institute for Computational Mechanics, Technical University of Munich, Boltzmannstr. 15, 85748 Garching, Germany. Tel.: +49 89 28915300; fax: +49 89 28915301}
\ead{kronbichler@lnm.mw.tum.de}
\address[add1]{Institute for Computational Mechanics, Technical University of Munich,\\ Boltzmannstr. 15, 85748 Garching, Germany}

\begin{abstract}
The present work develops hybrid multigrid methods for high-order discontinuous Galerkin discretizations of elliptic problems, which are, for example, a key ingredient of incompressible flow solvers in the field of computational fluid dynamics. Fast matrix-free operator evaluation on tensor product elements is used to devise a computationally efficient PDE solver. The multigrid hierarchy exploits all possibilities of geometric, polynomial, and algebraic coarsening, targeting engineering applications on complex geometries. Additionally, a transfer from discontinuous to continuous function spaces is performed within the multigrid hierarchy. This does not only further reduce the problem size of the coarse-grid problem, but also leads to a discretization most suitable for state-of-the-art algebraic multigrid methods applied as coarse-grid solver. The relevant design choices regarding the selection of optimal multigrid coarsening strategies among the various possibilities are discussed with the metric of computational costs as the driving force for algorithmic selections. We find that a transfer to a continuous function space at highest polynomial degree (or on the finest mesh), followed by polynomial and geometric coarsening, shows the best overall performance. The success of this particular multigrid strategy is due to a significant reduction in iteration counts as compared to a transfer from discontinuous to continuous function spaces at lowest polynomial degree (or on the coarsest mesh). The coarsening strategy with transfer to a continuous function space on the finest level leads to a multigrid algorithm that is robust with respect to the penalty parameter of the symmetric interior penalty method. Detailed numerical investigations are conducted for a series of examples ranging from academic test cases to more complex, practically relevant geometries. Performance comparisons to state-of-the-art methods from the literature demonstrate the versatility and computational efficiency of the proposed multigrid algorithms.
\end{abstract}

\begin{keyword}
discontinuous Galerkin method, high-order discretizations, interior penalty method, matrix-free algorithms, multigrid, time-to-solution
\end{keyword}

\end{frontmatter}


\section{Motivation}\label{Motivation}
Computer hardware progress towards high Flop-to-Byte ratios renders the data movement the deciding factor for efficient PDE solvers, especially for multigrid algorithms as their main algorithmic component in the case of elliptic operators. A consequence of this development is a stronger coupling of the numerical linear algebra part and computer science part in solver development, since black-box matrix-based solvers that are severly memory-bound are no longer competitive. Therefore, optimal multigrid solvers can not be designed with a view on convergence rates and iteration counts only, but need to take into account the implementation technique from the very beginning. With the goal in mind of optimizing time-to-solution, we believe that addressing these interdisciplinary challenges deserves special attention.

\begin{figure}[!ht]
\centering
\includegraphics[width=0.75\textwidth]{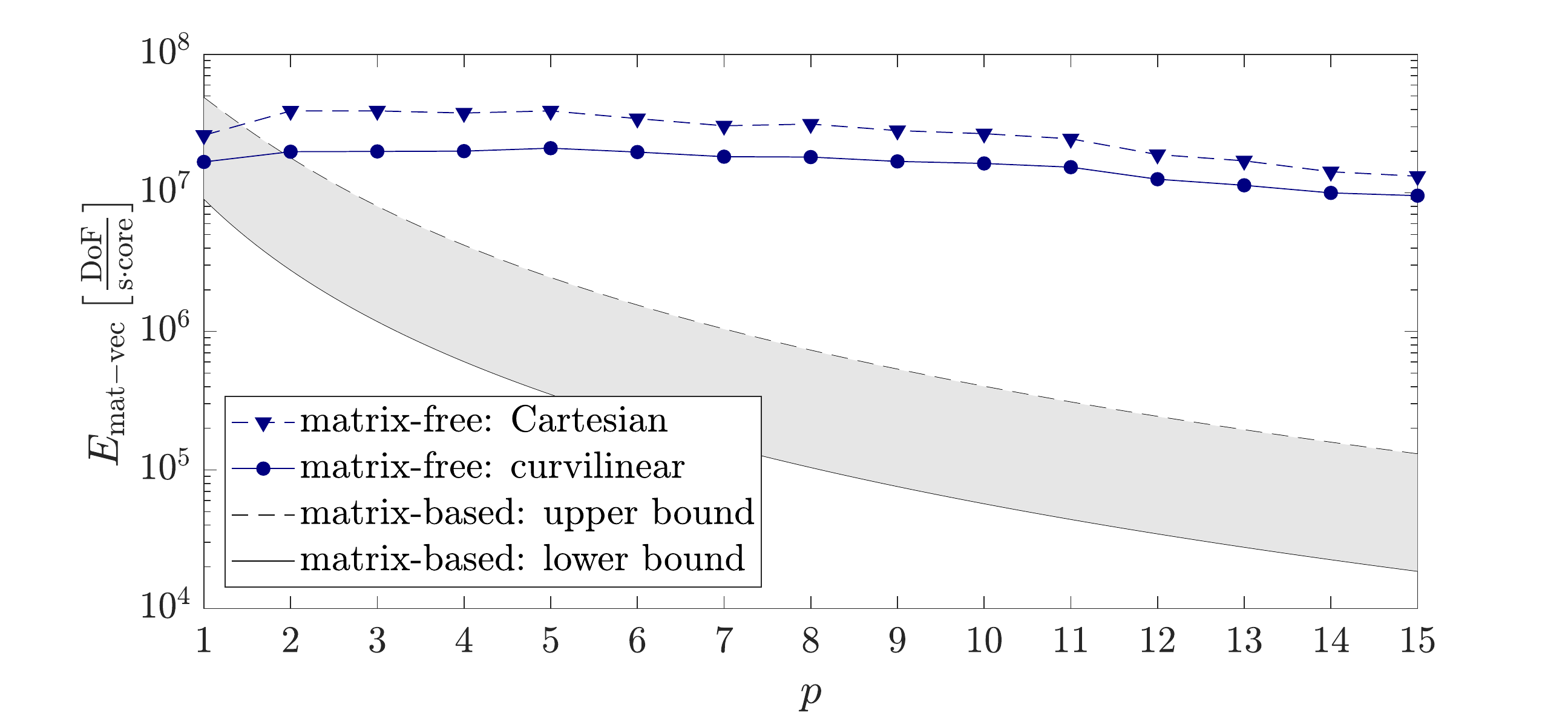}
\caption{Throughput of matrix-free versus matrix-based evaluation of scalar Laplace operator~$- \nabla^2 u$ discretized on a hexahedral mesh (3D). The throughput is measured in degrees of freedom (unknowns) processed per second per CPU core for one forward application of the discretized operator (matrix-vector product). The matrix-free version shows the measured performance for an interior penalty DG discretization with nodal Gauss--Lobatto basis where the experiments have been conducted on an Intel Skylake architecture (see Table~\ref{tab:Hardware} for details) for both a Cartesian mesh and a curvilinear mesh. The throughput measurements are done on one fully loaded compute node with the problem size ($25 \mathrm{MDoF}- 75 \mathrm{MDoF}$) being large enough to saturate caches. For the matrix-based implementation (considering only the matrix-vector product and neglecting assembly costs), theoretical lower and upper bounds are shown for the same Intel Skylake architecture, both assuming an optimal implementation. Depending on the chosen DG basis and the stencil-width (sparsity of block-matrices of neighboring elements), the matrix-based version will perform closer to the upper bound ($(p+1)^{2d}$ nonzeros in matrix, only block-diagonal taken into account, block matrices of neighboring elements are assumed sparse and are neglected) or lower bound ($(2d+1)(p+1)^{2d}$ nonzeros with the block-matrices of neighboring elements being dense). The implementation of the matrix-based variant is assumed optimal, i.e., the code is assumed to run with full memory throughput at the STREAM bandwidth of the hardware where only the non-zeros of the matrix as well as the input and output vectors have to be read from/written to memory. Furthermore, optimal spatial and temporal data locality is assumed, i.e., perfect utilization of cache-lines and caches.}
\label{fig:throuhgput_matrix_free_vs_matrix_based}
\end{figure}

\subsection{Matrix-free implementations and recent trends in computer hardware}\label{sec:ComputationalComplexities}
It is well known from the spectral element community that computationally efficient implementations of high-order finite element discretizations on tensor-product elements rely on matrix-free algorithms using the sum-factorization technique, see for example~\cite{Orszag1980, Kopriva2009, Deville2002, Karniadakis2013}. In the emerging field of high-order discontinuous Galerkin discretizations, recent implementations targeting current cache-based multicore architectures with high Flop-to-Byte ratios due to wide SIMD (single-instruction-multiple-data) units have been developed in~\cite{Kronbichler2012, Kronbichler2019fast, Muething2017arxiv}. These matrix-free evaluation routines using sum-factorization have a complexity of~$\mathcal{O}(p^{d+1})$ operations and~$\mathcal{O}(p^{d})$ data transfer from memory per element for polynomial degree~$p$ in~$d$ space dimensions. Traditional implementation strategies are based on the assembly of sparse matrices in the finite element discretization stage of the numerical algorithm and are handed over to the linear algebra part of the solver that can be applied in a black-box fashion to the discrete problem. However, due to increased complexity in terms of arithmetic operations and data transfer from main memory with complexity~$\mathcal{O}(p^{2d})$ for the matrix-vector product (and worse complexity for the assembly part), it is clear that these sparse matrix methods can not be competitive for high polynomial degrees. Initial studies~\cite{Vos2010, Cantwell2011} identified a break-even point of~$p \approx 5$ for~$d=3$ (and larger~$p$ for~$d=2$) beyond which a matrix-free approach with sum-factorization is faster. However, on modern computer hardware characterized by Flop-to-Byte ratios significantly larger than one, matrix-free algorithms with sum-factorization on hexahedral elements outperform sparse matrices already for polynomial degree~$p=2$, see~\cite{Kronbichler2012, Kronbichler2019}. This is also shown in Figure~\ref{fig:throuhgput_matrix_free_vs_matrix_based}, which compares the throughput measured in unknowns processed per second for the evaluation of the scalar Laplace operator in three space dimensions,~$d=3$, suggesting a break-even point of~$p=1-2$. Due to the paradigm shift in computer hardware, matrix-free algorithms nowadays tend to become memory-bound on recent hardware~\cite{Kronbichler2019fast} if implemented with a minimum of floating point operations and if vectorization is used. As a consequence of this trend, also solution techniques such as the hybridizable discontinuous Galerkin (HDG) method, see for example~\cite{Kirby2012, Yakovlev2016}, which reduces the global problem to the degrees of freedom on the mesh skeleton by elimination of inner degrees of freedom, can not keep up with fast matrix-free operator evaluation for quadrilateral/hexahedral elements on current computer hardware as investigated in detail in the recent article~\cite{Kronbichler2018}. In terms of Figure~\ref{fig:throuhgput_matrix_free_vs_matrix_based}, the gap would open at a slower pace for the HDG case, but still be more than one order of magnitude.

\subsection{Multigrid for high-order discretizations: State-of-the-art}
When it comes to the solution of (non-)linear systems of equations by means of iterative solution techniques, the evaluation of nonlinear residuals as well as the application of linear(ized) operators in Krylov solvers are readily available in a matrix-free implementation environment~\cite{Brown2010}. More importantly, however, also preconditioners should rely on matrix-free algorithms with optimal complexity, since the whole solver will otherwise run into similar bottlenecks with overwhelming memory transfer. Optimal complexity of all solver components is crucial in order to render high-order discretization methods more efficient in under-resolved application scenarios~\cite{Fehn2018b}. For elliptic PDEs, multigrid methods~\cite{Trottenberg2001} are among the most efficient solution techniques~\cite{Gholami2016}, especially because of their applicability to problems on complex geometries. In the context of high-order finite element discretizations, multigrid methods can be categorized into~$h$-multigrid,~$p$-multigrid, and algebraic multigrid (AMG) techniques. 

Geometric or~$h$-multigrid methods rely on a hierarchy of meshes. Robust $h$-multigrid techniques for high-order DG discretizations have been developed and analyzed in~\cite{Gopalakrishnan2003, Hemker2003, Brenner2005, Brenner2009} for uniformly refined meshes and in~\cite{Kanschat2004, Kanschat2008} for adaptively refined meshes. 
 Recent improvements of these algorithms towards high-performance, matrix-free implementations have been developed in~\cite{Kronbichler2018}, where a performance comparison of high-order continuous and discontinuous Galerkin discretizations as well as hybridizable discontinuous Galerkin methods is carried out. A GPU variant for continuous finite elements has been proposed in~\cite{Kronbichler2019}. The parallel efficiency for adaptively refined meshes is discussed in~\cite{Clevenger2019}. Large-scale applications of these~$h$-multigrid methods can be found in the field of computational fluid dynamics (CFD) and the incompressible Navier--Stokes equations~\cite{Fehn2018b, Krank2017}. 

For spectral element discretizations,~$p$-multigrid, or synonymously spectral element multigrid, is frequently used, where the multigrid hierarchy is obtained by reducing the polynomial degree of the shape functions. Coarsening and multigrid transfer is particularly simple since the function spaces of different multigrid levels are nested also for element geometries deformed via high-order mappings and all operations are element-local. This approach has first been proposed and theoretically analyzed in~\cite{Ronquist1987,Maday1988}, and later investigated, for example, in~\cite{Helenbrook2003, Helenbrook2008, Mascarenhas2010, Lottes2005, Stiller2017, Stiller2016, Stiller2017b, Huismann2019}. A related hierarchical scale separation solver is proposed in~\cite{Aizinger2015}. Polynomial multigrid techniques are also frequently used to solve the compressible Euler equations~\cite{Rasetarinera2001, Bassi2003, Fidkowski2004, Nastase2006JCP, Luo2006, Hillewaert2006, Mascarenhas2009, Bassi2009, Helenbrook2016} and compressible Navier--Stokes equations~\cite{Fidkowski2005, Persson2008, Shahbazi2009, Diosady2009, Bassi2011, Luo2012, Ghidoni2014}. 

Algebraic multigrid techniques extract all information from the assembled system matrix to generate coarser levels and are attractive as they can be applied in a black-box fashion. AMG is considered in~\cite{Heys2005} for high-order continuous Galerkin discretizations and in~\cite{Lasser2001, Prill2009, Olson2011} for (high-order) discontinuous Galerkin discretizations. AMG applied directly to high-order DG discretizations faces several challenges, among them the construction of robust smoothers for matrices that lose diagonal dominance, but most importantly the computational complexity of matrix-based algorithms (especially for three-dimensional problems) compared to their matrix-free counterparts, see Figure~\ref{fig:throuhgput_matrix_free_vs_matrix_based}. These limitations are also exemplified by the fact that AMG methods for high-order discretizations have only been applied to small two-dimensional problems in the works mentioned above. For reasons of computational complexity (see Section~\ref{sec:ComputationalComplexities}), it appears to be inevitable to combine algebraic multigrid techniques with some form of geometric coarsening to achieve a computationally efficient approach for practical applications~\cite{Prill2009,Bastian2012, Siefert2014}. 

Multigrid transfer operators are typically negligible in terms of computational costs when implemented in a matrix-free way with optimal complexity~\cite{Kronbichler2019, Bastian2019}. Therefore, multigrid smoothers and coarse-grid solvers remain as the main performance-relevant multigrid components. Adhering to the matrix-free paradigm poses two major challenges that need to be addressed and further improved:

\begin{figure}[!ht]
\centering
  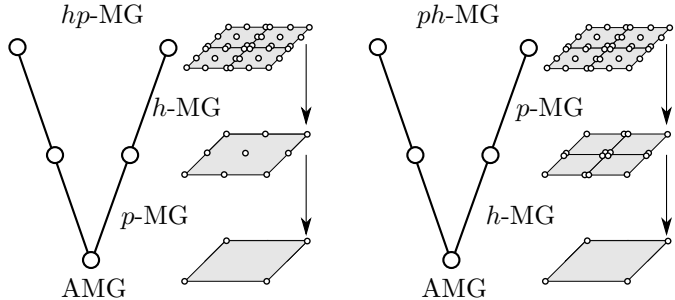
\caption{Illustration of combined geometric--polynomial--algebraic multigrid algorithms for nodal high-order discontinuous Galerkin discretizations.}
\label{fig:HybridMultigrid}
\end{figure}

\begin{itemize}
\item Matrix-free smoothers: To fully exploit the advantages of matrix-free algorithms with sum-factorization for operator evaluation, multigrid smoothers should make use of these algorithms as well, but this problem has so far only been solved for certain types of smoothers (and mainly for elliptic problems). Polynomial smoothers based on the Chebyshev iteration~\cite{Adams2003,Sundar2015} or explicit Runge--Kutta methods~\cite{Helenbrook2003,Luo2006,Hillewaert2006,Luo2012,RuedaRamirez2019} can be immediately realized in a matrix-free way, are inherently parallel, and are therefore widely used in a high-order context. More challenging are smoothers based on elementwise inversion of operators such as block Jacobi, block Gauss--Seidel, block ILU, or (overlapping) Schwarz smoothers. Many implementations rely on matrix-based algorithms for smoothers~\cite{Rasetarinera2001, Fidkowski2004, Fidkowski2005, Nastase2006JCP,Bassi2009, Persson2008,Shahbazi2009,Diosady2009,Mascarenhas2009,Bassi2011,Ghidoni2014,Helenbrook2016}, limiting applicability mainly to two-dimensional problems, while three-dimensional problems become prohibitively expensive for higher-polynomial degrees due to the complexity of~$\mathcal{O}(p^{2d+1})$ for the assembly,~$\mathcal{O}(p^{3d})$ for factorizations, and~$\mathcal{O}(p^{2d})$ for matrix-vector products. On Cartesian meshes and tensor product elements, elementwise inversion of operators with optimal complexity is possible via the fast diagonalization method~\cite{Lynch1964}, which has first been applied in~\cite{Couzy1994, Couzy1995} in the context of spectral element discretizations and analyzed in more detail in~\cite{Fischer2000, Lottes2005, Fischer2005, Stiller2017, Stiller2016, Stiller2017b} in the context of overlapping Schwarz preconditioners and multigrid smoothers. Other tensor-product based preconditioners for high-order DG discretizations that exploit fast matrix-free operator evaluation with sum-factorization and that are applicable to more complex operators (convection--diffusion, Navier--Stokes) and non-Cartesian meshes have been proposed recently in~\cite{Bastian2019, Pazner2018}, suggesting that the complexity can be reduced to~$\mathcal{O}(p^{d+2})$ in a general setting.

\item Hybrid multigrid algorithms: Due to the improved efficiency of matrix-free evaluation routines and better parallel scalability~\cite{Sundar2012} as compared to sparse matrix-vector products, the generation of coarser multigrid levels should rely on non-algebraic coarsening, i.e., mesh (or geometric) coarsening and reducing the polynomial degree of the function space for higher order methods, leading to the idea of hybrid~$hp$- and~$ph$-multigrid methods as illustrated in Figure~\ref{fig:HybridMultigrid}. It is beneficial to stay as long as possible on the matrix-free curve in Figure~\ref{fig:throuhgput_matrix_free_vs_matrix_based} and to exploit all possibilities of geometric and polynomial coarsening in the multigrid hierarchy. For complex engineering applications, the number of geometric mesh levels is low (typically~$0,1,2$) and coarse meshes might consist of millions of elements so that simple iterative solution techniques like a conjugate gradient iteration with Jacobi preconditioner used as coarse-grid solver would become too expensive and dominate the overall computational costs of the multigrid solver. Applying algebraic multigrid techniques for the coarse-grid problem discretized with linear finite elements is a good compromise between effectiveness of coarse-grid preconditioning and computational efficiency, since in this regime sparse matrix-vectors products are competitive to matrix-free evaluation routines, see Figure~\ref{fig:throuhgput_matrix_free_vs_matrix_based}. At the same time, smoothers for algebraic multigrid methods work best at low polynomial degrees due to a better diagonal dominance of the matrix as opposed to high-order shape functions~\cite{Kronbichler2018}. In Table~\ref{tab:HybridMultigridApproaches}, we summarize important contributions in the field of hybrid multigrid solvers. In~\cite{Helenbrook2003, Helenbrook2001}, hybrid multigrid solvers combining~$p$-MG with~$h$-MG have been presented for high-order discretizations. In~\cite{Nastase2006JCP,Shahbazi2009},~$p$-multigrid is used along with algebraic multigrid for the coarse problem. In terms of~$p$-multigrid, the works~\cite{Dobrev2006, Bastian2012, Siefert2014} can be categorized as two-level algorithms with transfer to continuous linear shape functions from the fine level to the coarse level, which is solved by an~$h$-multigrid approach in~\cite{Dobrev2006} and an algebraic multigrid approach in~\cite{Bastian2012, Siefert2014}. These works have the limitation that the underlying implementation is not matrix-free and, therefore, suffer from the complexity of matrix-based approaches. The main limitation of the approach in~\cite{Sundar2012} for hexahedral meshes based on the octree concept is that it only supports linear continuous finite elements (a similar approach for tetrahedral elements is presented in~\cite{Lu2014}). An extension towards~$p$-multigrid has been done in~\cite{Rudi2015} but results are limited to linear and quadratic shape functions. The approach in~\cite{OMalley2017} combines~$p$-multigrid with AMG but uses expensive matrix-based implementations which could explain why results are only shown for quadratic elements. In more recent works, hybrid multigrid algorithms for high-order methods with completely matrix-free implementation are discussed in~\cite{Bastian2019}, extending a previous work~\cite{Bastian2012} towards a matrix-free implementation developed in~\cite{Muething2017arxiv, Kempf2018arxiv} and available in the EXADUNE project~\cite{Exadune2014}. That work does not exploit geometric coarsening ($h$-multigrid) and the high-order discretization with degree~$p$ is immediately reduced to a linear space within the multigrid algorithm (and is therefore categorized as a two-level algorithm rather than~$p$-multigrid). Algebraic multigrid is employed for the coarse problem. An elaborate matrix-free implementation in the context of $h$-multigrid solvers is presented in~\cite{Kronbichler2018} based on the matrix-free implementation developed in~\cite{Kronbichler2012,Kronbichler2019fast} and available in the~\texttt{deal.II} finite element library~\cite{dealII90}. That work clearly improves the performance numbers of sophisticated geometric multigrid solvers shown in~\cite{Gholami2016}. One drawback of this pure~$h$-multigrid approach is that its applicability is limited to problems where the coarse grid is comparably simple. Hybrid multigrid techniques in the context of HDG discretizations are considered, e.g., in~\cite{Fabien2019}.
\end{itemize}

\begin{table}[t]
\caption{Hybrid multigrid algorithms: relevant publications from the literature addressing combined $h$-,~$p$-, and algebraic multigrid methods are categorized in terms of high-order discretizations ($p > 2$), matrix-free implementations, $h$-MG, $p$-MG, and AMG. Legend: \Checkmark $=$ fulfilled,(\Checkmark) $=$ partly fulfilled, \XSolidBrush $=$ not fulfilled. The category~$p$-MG is partly fulfilled (\Checkmark) if a two-level algorithm is considered with transfer from high-order space of degree~$p$ directly to linear space with~$p=1$.}
\label{tab:HybridMultigridApproaches}
\renewcommand{\arraystretch}{1.1}
\begin{center}
\begin{tabular}{lccccc}
\hline
Study & high-order & matrix-free & $h$-MG & $p$-MG & AMG\\
\hline
Helenbrook et al.~\cite{Helenbrook2003}  & \Checkmark & \XSolidBrush &  \Checkmark &  \Checkmark & \XSolidBrush\\
Nastase et al.~\cite{Nastase2006JCP}, Shahbazi et al.~\cite{Shahbazi2009} & \Checkmark & \XSolidBrush &  \XSolidBrush &  \Checkmark & \Checkmark\\
Dobrev et al.~\cite{Dobrev2006} 	 & \XSolidBrush & \XSolidBrush   &   \Checkmark &  (\Checkmark) & \XSolidBrush\\
Bastian et al.~\cite{Bastian2012}, Siefert et al.~\cite{Siefert2014} 	 & \Checkmark & \XSolidBrush   &   \XSolidBrush &  (\Checkmark) & \Checkmark\\
Sundar et al.~\cite{Sundar2012} & \XSolidBrush & \Checkmark &  \Checkmark & \XSolidBrush & \Checkmark\\
Lu et al.~\cite{Lu2014} & \XSolidBrush & \XSolidBrush &  \Checkmark & \XSolidBrush & \Checkmark\\
Rudi et al.~\cite{Rudi2015} & \XSolidBrush & \Checkmark &  \Checkmark & \Checkmark & \Checkmark\\
O'Malley et al.~\cite{OMalley2017} & \XSolidBrush & \XSolidBrush &  \XSolidBrush &  \Checkmark & \Checkmark\\
Bastian et al.~\cite{Bastian2019} 	 & \Checkmark & \Checkmark   &   \XSolidBrush &  (\Checkmark) & \Checkmark\\
Kronbichler and Wall~\cite{Kronbichler2018} & \Checkmark & \Checkmark   &   \Checkmark &  \XSolidBrush & \XSolidBrush\\
present work 							& \Checkmark & \Checkmark   &   \Checkmark &  \Checkmark & \Checkmark\\
\hline
\end{tabular}
\end{center}
\renewcommand{\arraystretch}{1}
\end{table}

There exist other techniques as well with the aim to overcome the complexity of matrix-based methods for high polynomial degrees. Preconditioners and multigrid methods applied to a low-order re-discretization of the operator on a mesh with vertices located on the nodes of the high-order discretization is a well-known technique originating from~\cite{Deville1985, Deville1990} and has been analyzed for example in~\cite{Fischer1997, Lottes2005, Heys2005, Brown2010, Sundar2015, Pazner2019arxiv}. Such an approach is not considered here.

\subsection{Contributions of the present work}
The present work extends our previous work in~\cite{Kronbichler2018} towards hybrid multigrid techniques combining geometric ($h$), polynomial ($p$), and algebraic coarsening. Our goal is to devise a multigrid solver applicable to engineering problems with complex geometry characterized by coarse grids with many elements. As can be seen from Table~\ref{tab:HybridMultigridApproaches}, the individual components relevant for efficient hybrid multigrid methods are covered by different works. However, none of these works fulfills all properties and it is the aim of the present work to fill this gap.

As a model problem, the constant-coefficient Poisson equation in three space dimensions is studied in this work. With respect to the choice of multigrid smoothers, this study makes use of Chebyshev accelerated Jacobi smoothers which have the characteristic that convergence rates are independent of~$h$ and mildly dependent on~$p$, see~\cite{Kronbichler2018, Sundar2015}. Chebyshev smoothing is particularly attractive since it only requires application of the matrix-vector product and the inverse diagonal of the system matrix, i.e., the smoother benefits from fast matrix-free evaluation routines and is efficient in a parallel setting. Although more aggressive smoothers based on overlapping Schwarz methods resulting in lower iteration counts exist, it should be noted that Chebyshev smoothing is nonetheless highly efficient and comparative studies would need to be carried out to answer which approach is more efficient, see the initial investigation in~\cite{Kronbichler2019arxiv}. These aspects are, however, beyond the scope of the present study.

In case of discontinuous Galerkin discretizations, a transfer from discontinuous to continuous function spaces (denoted as DG-to-FE transfer) should be considered in addition to~$h$- and~$p$-coarsening in order to further reduce the size of the coarse-grid problem. For example, the problem size is reduced by a factor of~$2^d$ for linear elements with~$p=1$. Moreover, this approach is also beneficial in order to reduce iteration counts for the coarse-grid problem, due to the fact that existing AMG implementations and smoothers are often most effective on continuous function spaces. However, it is unclear whether to perform the DG-to-FE transfer on the high-order polynomial space~$p$ or for the coarse problem at~$p=1$, or likewise on the finest mesh or the coarsest mesh. 
It is a main finding of the present work that a DG-to-FE transfer at the fine level is beneficial, both in terms of iteration counts and overall computational costs. Furthermore, we demonstrate that this approach results in a multigrid algorithm whose convergence rates are independent of the interior penalty factor. This leads to multigrid coarsening strategies denoted as~$chp$- or~$cph$-multigrid, with a transfer to continuous~($c$) function spaces on the finest level followed by geometric~($h$) and polynomial~($p$) coarsening before the coarse-grid solver (e.g., AMG) is invoked.

In summary, the present work discusses the relevant design choices in the context of hybrid multigrid algorithms, i.e., combined geometric--polynomial--algebraic multigrid techniques, with an emphasis on computational costs as the driving force for algorithmic selections. The performance of these methods is detailed using a state-of-the-art matrix-free implementation, considering a series of increasingly complex problems.

\subsection{Outline} The model problem studied in this work and the discontinuous Galerkin discretization are introduced in Section~\ref{sec:ModelProblem}. Section~\ref{sec:HybridMultigrid} discusses the hybrid multigrid algorithm including the main multigrid components such as smoothers, coarsening strategies and transfer operators, as well as the coarse-level solver. The matrix-free implementation is summarized in Section~\ref{sec:MatrixFree} which is the key to an efficient hybrid multigrid solver. Numerical results are shown in Section~\ref{sec:Results}, and we conclude in Section~\ref{sec:Summary} with a summary of our results and an outlook on future work.

\section{High-order discontinuous Galerkin discretization of the Poisson equation}\label{sec:ModelProblem}
As a model problem, we consider the Poisson equation discretized by discontinuous Galerkin methods with a focus on high-order polynomial spaces. Let us briefly motivate the use of discontinuous Galerkin discretizations for the Poisson problem. While continuous finite element discretizations might be regarded a suitable discretization scheme as well due to a reduced number of unknowns, DG discretizations can have an advantage over continuous discretizations for non-smooth problems in terms of accuracy versus computational costs due to a better approximation in proximity to a singularity~\cite{Kronbichler2018}. Furthermore, DG discretizations of the Poisson equation arise naturally from certain model problems such as the incompressible Navier--Stokes equations discretized with discontinuous Galerkin methods. For this type of problems, efficient multigrid methods for Poisson problems are a key ingredient determining overall efficiency. Large-scale applications in the context of incompressible turbulent flows can be found in~\cite{Krank2017, Fehn2018b}, for earth mantle convection problems (with variable coefficients) in~\cite{Rudi2015}, or for porous media flow in~\cite{Bastian2014}. The constant coefficient Poisson equation reads
\begin{align*}
- \nabla^2 u = f \; \text{in}  \; \Omega \in \mathbb{R}^d \; .
\end{align*}
On the domain boundary~$\Gamma = \partial \Omega$, Dirichlet boundary conditions,~$u=g$ on~$\Gamma_{\mathrm{D}}$, and Neumann boundary conditions,~$\Grad{u}\cdot \bm{n} = h$ on~$\Gamma_{\mathrm{N}}$, are prescribed, with~$\Gamma_{\mathrm{D}} \cup \Gamma_{\mathrm{N}} = \Gamma$ and~$\Gamma_{\mathrm{D}} \cap \Gamma_{\mathrm{N}} = \emptyset$.

We consider meshes composed of hexahedral elements~$\Omega_e$,~$e=1, ...,N_{\text{el}}$, that may be arbitrarily deformed via a high-order polynomial mapping~$\bm{x}\left(\bm{\xi}\right)$ from the reference element~$\hat{\Omega}_e=[0,1]^d$ with coordinates~$\bm{\xi}$ to the physical element~$\Omega_{e}$ with coordinates~$\bm{x}$. The space of test and trial functions is given as
\begin{align}
\mathcal{V}_{h} &= \left\lbrace u_h\in L_2(\Omega_h)\; : \; u_h\left(\bm{x}(\boldsymbol{\xi})\right)\vert_{\Omega_{e}} = \hat{u}_h^e(\boldsymbol{\xi})\vert_{\hat{\Omega}_{e}}\in \mathcal{V}_{h,e}=\mathcal{Q}_{p}(\hat{\Omega}_{e})\; ,\;\; \forall e=1,\ldots,N_{\text{el}} \right\rbrace\; ,
\end{align}
Here,~$\mathcal{Q}_{p}(\hat{\Omega}_{e})$ denotes the space of polynomials of tensor degree~$\leq p$ defined on the reference element~$\hat{\Omega}_e$, i.e.,
\begin{align}
\hat{u}_h^e(\boldsymbol{\xi},t) = \sum_{i_1,...,i_d=0}^{p} N_{i_1...i_d}^{p}(\boldsymbol{\xi})u_{i_1...i_d}^e(t)\; .
\end{align}
The multidimensional shape functions are formed by a tensor product of one-dimensional shape functions,~$N_{i_1...i_d}^{p}(\boldsymbol{\xi})=\prod_{n=1}^{d} l_{i_n}^p(\xi_n)$. Lagrange polynomials~$l_i^p(\xi)$ of degree~$p$ based on the Legendre--Gauss--Lobatto nodes are a common choice for a nodal basis due to their conditioning and are therefore considered in the present work. As usual, volume and face integrals in the weak formulation are computed by means of Gaussian quadrature with~$p+1$ points per coordinate direction, ensuring exact integration on affine geometries. Note that the tensor product structure of both the shape functions and the quadrature rule is important in order to use fast matrix-free evaluation routines exploiting sum-factorization.

The weak formulation of the Poisson problem written in primal formulation reads: Find~$u_h\in \mathcal{V}_h$ such that
\begin{align}
a_{h}^{e}\left(v_h,u_h \right) = \intele{v_h}{f} \;\; \forall v_h \in \mathcal{V}_{h,e}\; , e=1,...,N_{\text{el}} \; ,\label{eq:WeakFormPoissonEquation}
\end{align}
where
\begin{align}
a_{h}^{e}\left(v_h,u_h \right) = \intele{\Grad{v_h}}{\Grad{u_h}}-\inteleface{\Grad{v_h}}{\left(u_h-u_h^*\right)\bm{n}} - \inteleface{v_h}{\bm{\sigma}_h^*\cdot\bm{n}} \; .\label{eq:WeakFormDG}
\end{align}
We use the notation~$\intele{v}{u} = \int_{\Omega_e} v \odot u \; \mathrm{d}\Omega$ and~$\inteleface{v}{u} = \int_{\partial \Omega_e} v \odot u \; \mathrm{d} \Gamma$ for volume and face integrals, respectively, with inner products symbolized by~$\odot$. As an important representative of the class of DG discretization methods we consider the symmetric interior penalty method~\cite{Arnold1982,Arnold2002} with numerical fluxes
\begin{align}
u_h^* &= \avg{u_h}\; ,\label{eq:InteriorPenaltyFlux1}\\
\bm{\sigma}_h^* &= \avg{\Grad{u}_h}-\tau \jump{u_h}\; ,\label{eq:InteriorPenaltyFlux2}
\end{align}
where~$\avg{u_h} = \left( u_h^- + (u_h)^+\right)/2$ is the average operator and~$\jump{u_h} = u_h^- \otimes \bm{n}^- + u_h^+ \otimes \bm{n}^+$ the jump operator, and~$(\cdot)^-$,~$(\cdot)^+$ denote two neighboring elements~$e^-$,~$e^+$. Boundary conditions are imposed weakly via the mirror principle~\cite{Hesthaven2007}, setting
\begin{align}
u_h^+ = \begin{cases}
-u_h^- + 2g & \text{on} \, \Gamma_{h,\mathrm{D}}\; ,\\
u_h^- & \text{on} \, \Gamma_{h,\mathrm{N}}\; ,
\end{cases}
\hspace{0,5cm}\text{and}\hspace{0,5cm}
\Grad{u_h^+}\cdot \bm{n} = \begin{cases}
\Grad{u_h^-}\cdot \bm{n} & \text{on} \, \Gamma_{h,\mathrm{D}}\; ,\\
- \Grad{u_h^-}\cdot \bm{n} + 2h & \text{on} \, \Gamma_{h,\mathrm{N}}\; .
\end{cases}\label{eq:BoundaryConditions}
\end{align}
Inserting equation~\eqref{eq:BoundaryConditions} into equation~\eqref{eq:WeakFormDG}, the weak form can be split into homogeneous and inhomogeneous contributions,~$a_{h}^{e}\left(v_h,u_h, g, h\right) = a_{h,\mathrm{hom}}^{e}\left(v_h,u_h \right) + a_{h,\mathrm{inhom}}^{e}\left(v_h, g, h\right)$.
The definition of the penalty parameter~$\tau_e$ according to~\cite{Hillewaert2013} for quadrilateral/hexahedral elements is used
\begin{align}
\tau_e = (p+1)^2 \frac{A\left(\partial \Omega_e \setminus \Gamma_h\right)/2 + A\left(\partial \Omega_e \cap \Gamma_h\right)}{V\left(\Omega_e\right)}\; ,
\end{align}
with the element volume~$V\left(\Omega_e\right) = \int_{\Omega_e}\mathrm{d}\Omega$ and the surface area~$A(f) = \int_{f\subset\partial\Omega_e}\mathrm{d}\Gamma$. The penalty parameter~$\tau$ in equation~\eqref{eq:InteriorPenaltyFlux2} is given as~$\tau = \max\left(\tau_{e^-},\tau_{e^+}\right)$ on interior faces~$f \subseteq \partial \Omega_e \setminus \Gamma_h$, and~$\tau = \tau_e$ on boundary faces~$f \subseteq \partial \Omega_e \cap \Gamma_h$.

For the multigrid algorithm detailed below, coarser discretizations of the Laplace operator are required, which is realized by evaluating the operator for modified discretization parameters~$h$ and~$p$ (including the interior penalty parameter), i.e., on a coarser mesh or for a lower polynomial degree~$p$. In the literature, this approach is sometimes denoted as re-discretization, as opposed to a Galerkin product. Further, a transfer from discontinuous to continuous function spaces is considered in our hybrid multigrid algorithm. The conforming finite element (FE) space is given as
\begin{align}
\mathcal{V}_{h}^{\mathrm{FE}} &= \left\lbrace u_h\in H^1(\Omega_h)\; : \; u_h\left(\bm{x}(\boldsymbol{\xi})\right)\vert_{\Omega_{e}} = \hat{u}_h^e(\boldsymbol{\xi})\vert_{\hat{\Omega}_{e}}\in \mathcal{V}_{h,e}=\mathcal{Q}_{p}(\hat{\Omega}_{e})\; ,\;\; \forall e=1,\ldots,N_{\text{el}} \right\rbrace\; ,
\end{align}
and the weak form of the (negative) Laplace operator simplifies to
\begin{align}
a_{h,\mathrm{FE}}^{e}\left(v_h,u_h \right) = \intele{\Grad{v_h}}{\Grad{u_h}} \; .\label{eq:WeakFormFE}
\end{align}
Dirichlet boundary conditions are imposed strongly for the FE discretization, but only the homogeneous operator is required inside the multigrid algorithm. When assembling the coarse-level matrix for the AMG coarse solver, constrained degrees of freedom are kept in the system with diagonal entries set to 1 to ensure that the matrix is invertible.

In matrix-vector notation, the discrete problem can be written as the linear system of equations
\begin{align}
\bm{A} \bm{u} = \bm{b} \; ,
\end{align}
where~$\bm{A} \in \mathbb{R}^{N \times N}$,~$\bm{u},\bm{b}\in \mathbb{R}^{N}$ with the problem size (number of unknowns or degrees of freedom) denoted by~$N = N_{\mathrm{el}} (p+1)^d$.  Contributions from inhomogeneous boundary conditions are included in the right-hand side vector~$\bm{b}$ and the matrix~$\bm{A}$ only accounts for the homogeneous part~$a_{h,\mathrm{hom}}^{e}\left(v_h,u_h \right)$. The matrix~$\bm{A}$ is not assembled explicitly (apart from the algebraic multigrid coarse solver), since iterative solvers and multigrid smoothers only require the action of~$\bm{A}$ applied to a vector,~\footnote{A notation like~$\bm{a}(\bm{u})$ rather than~$\bm{A} \bm{u}$ would be more consistent in the matrix-free context, but we adhere to the matrix-vector notation as this is the common notation in linear algebra.} which is realized by means of fast matrix-free evaluations~(Section~\ref{sec:MatrixFree}).

\section{Hybrid multigrid solver}\label{sec:HybridMultigrid}

\begin{algorithm}[!ht]
\caption{Preconditioned conjugate gradient algorithm (solves~$\bm{A}\bm{x} = \bm{b}$ to given tolerance)}\label{alg:CG}
\begin{algorithmic}[1] 
\Function{SolverCG}{$\bm{A}, \bm{x}, \bm{b}$}
\State $\bm{r} = \bm{b} - \bm{A}\bm{x}$
\State $\Vert \bm{r}_0 \Vert = \Vert \bm{r} \Vert = \Call{Norm}{\bm{r}}$
\State $\bm{v} = \bm{M}^{-1} \bm{r}$ \Comment e.g., \Call{MultigridVCycle}{$L, \bm{A}, \textbf{0}, \bm{r}$}
\State $\bm{p} = \bm{v}$
\State $\delta = \bm{r}^T \bm{v}$
\While{$\Vert \bm{r} \Vert / \Vert \bm{r}_0 \Vert > \texttt{reltol} $ \textbf{and} $\Vert \bm{r} \Vert > \texttt{abstol}$}
\State $\bm{v} = \bm{A} \bm{p}$ \label{algCG:mat-vec}
\State $\omega = \delta/(\bm{p}^T \bm{v})$
\State $\bm{x} \gets \bm{x} + \omega \bm{p}$
\State $\bm{r} \gets \bm{r} - \omega \bm{v}$
\State $\Vert \bm{r} \Vert = \Call{Norm}{\bm{r}}$
\State $\bm{v} = \bm{M}^{-1} \bm{r}$ \Comment e.g., \Call{MultigridVCycle}{$L, \bm{A}, \textbf{0}, \bm{r}$} \label{algCG:preconditioner}
\State $\delta' = \delta$
\State $\delta = \bm{r}^T \bm{v}$
\State $\beta = \delta / \delta'$
\State $\bm{p} \gets \bm{v} + \beta \bm{p}$
\EndWhile
\EndFunction
\end{algorithmic}
\end{algorithm}

\begin{algorithm}[!ht]
\caption{Multigrid V-cycle (solves~$\bm{A}\bm{x} = \bm{b}$ approximately)}\label{alg:VCycle}
\begin{algorithmic}[1] 
\Function{MultigridVCycle}{$l, \bm{A}^{(l)}, \bm{x}^{(l)}, \bm{b}^{(l)}$}
\If {$l=0$}
\State $\bm{x}^{(0)} \gets$ \Call{CoarseLevelSolver}{$\bm{A}^{(0)}, \bm{x}^{(0)},\bm{b}^{(0)}$} \Comment coarse-level solver, e.g., AMG \label{algVCycle:coarse-grid-solver}
\State \Return $\bm{x}^{(0)}$
\Else
\State $\bm{x}^{(l)} \gets$ \Call{Smooth}{$\bm{A}^{(l)}, \bm{x}^{(l)}, \bm{b}^{(l)}, n_{\mathrm{s}}$} \Comment pre-smoothing \label{algVCycle:pre-smoothing}
\State $\bm{r}^{(l)} = \bm{b}^{(l)} - \bm{A}^{(l)} \bm{x}^{(l)}$ \Comment calculate residual \label{algVCycle:residual}
\State $\bm{b}^{(l-1)} = \bm{R}_{l}^{l-1}\bm{r}^{(l)}$\Comment restriction \label{algVCycle:restriction}
\State $\bm{x}^{(l-1)} \gets$ \Call{MultigridVCycle}{$l-1, \bm{A}^{(l-1)}, \textbf{0}, \bm{b}^{(l-1)}$} \Comment coarse-level correction \label{algVCycle:coarse-grid-correction}
\State $\bm{x}^{(l)} \gets \bm{x}^{(l)} + \bm{P}_{l-1}^{l}\bm{x}^{(l-1)}$ \Comment prolongation \label{algVCycle:prolongation}
\State $\bm{x}^{(l)} \gets$ \Call{Smooth}{$\bm{A}^{(l)}, \bm{x}^{(l)}, \bm{b}^{(l)}, n_{\mathrm{s}}$} \Comment post-smoothing  \label{algVCycle:post-smoothing}
\State \Return $\bm{x}^{(l)}$
\EndIf
\EndFunction
\end{algorithmic}
\end{algorithm}

\begin{algorithm}[!ht]
\caption{Chebyshev-accelerated Jacobi smoother (solves~$\bm{A}\bm{x} = \bm{b}$ approximately)}\label{alg:Chebyshev}
\begin{algorithmic}[1] 
\Function{ChebyshevSmoother}{$\bm{A}, \bm{x}_0, \bm{b}, n_{\mathrm{s}}$}
\For {$j=0,\ldots,n_{\mathrm{s}}-1$}
\State $\bm{x}_{j+1} = \bm{x}_j + \sigma_j \left(\bm{x}_j - \bm{x}_{j-1}\right) +  \theta_j \bm{D}^{-1} \left( \bm{b} - \bm{A} \bm{x}_j\right)$
\EndFor
\State \Return $\bm{x}_{n_{\mathrm{s}}}$
\EndFunction
\end{algorithmic}
\end{algorithm}

The basic multigrid idea is to tackle oscillatory errors by smoothing and to tackle smooth errors by coarse-grid corrections, which applies to all types of multigrid coarsening (geometric, polynomial, algebraic) discussed here. We use multigrid as a preconditioner inside a Krylov solver instead of using multigrid as a solver. This approach is sometimes also denoted as a Krylov-accelerated multigrid method. The combination of a multigrid cycle with a Krylov method can be expected to be more robust and to result in lower iteration counts in general as opposed to pure multigrid solvers, see~\cite{Lottes2005,Stiller2017,Stiller2016,Stiller2017b,Shahbazi2009,Sundar2015}, especially for anisotropic problems. Since it appears that this is the most frequent use case in practice, we follow this strategy in the present work. Some performance numbers could be improved by alternative multigrid flavors, e.g., by considering full multigrid cycles~\cite{Kronbichler2019}. The performance considerations and convergence
rates in this work also apply to full multigrid where only a single or only few iterations on the finest level are needed. Due to the symmetric positive definite nature of the model problem considered here, we use the conjugate gradient (CG) algorithm~\cite{Hestenes1952, Saad2003} as Krylov solver, which is detailed in Algorithm~\ref{alg:CG}. The algorithmic components which are of main interest are the application of the discretized operator in line~\ref{algCG:mat-vec} and the application of the preconditioner in line~\ref{algCG:preconditioner}. Other components are vector update operations and inner products (involving global communcation), but these parts of the algorithm are overall of subordinate importance since the computational costs are mainly determined by operator evaluation and the multigrid cycle called in the preconditioning step. However, it should be pointed out that the costs of all parts of the algorithm are explicitly taken into account by the experimental cost measures used in the present work, in the spirit of parallel textbook multigrid efficiency~\cite{ Gmeiner2015b}, see Section~\ref{sec:Metrics}.

In the preconditioning step of the conjugate gradient solver (preconditioner~$\bm{M}$), the operator~$\bm{A}$ is inverted approximately by performing one multigrid V-cycle according to Algorithm~\ref{alg:VCycle} with initial solution~$\bm{x}^{(L)} = \bm{0}$, where~$L$ denotes the finest level. Pre- and postsmoothing are done in lines~\ref{algVCycle:pre-smoothing} and~\ref{algVCycle:post-smoothing}, respectively, and the residual evaluation in line~\ref{algVCycle:residual}. The same number of smoothing steps~$n_{\mathrm{s}}$ is used for both pre- and postsmoothing and for all multigrid levels~$0 < l \leq L$. These steps typically form the most expensive part of the multigrid algorithm as long as the workload in degrees of freedom per core is sufficiently large, i.e., away from the strong-scaling limit where latency effects become dominant. The coarse-level correction is called in line~\ref{algVCycle:coarse-grid-correction}, recursively calling the multigrid V-cycle for the next coarser level~$l-1$ until the coarsest level~$l=0$ is reached, on which the coarse-level solver is called (line~\ref{algVCycle:coarse-grid-solver}). Restriction (operator~$\bm{R}_{l}^{l-1}$) and prolongation (operator$\bm{P}_{l-1}^{l}$) are done in lines~\ref{algVCycle:restriction} and~\ref{algVCycle:prolongation}, respectively.

\subsection{Chebyshev-accelerated Jacobi smoother}

In the context of matrix-free methods analyzed here, an attractive multigrid smoother is a Chebyshev-accelerated Jacobi smoother~\cite{Adams2003}, which requires the diagonal~$\bm{D}$ of the operator~$\bm{A}$ as well as the application of the matrix-vector product~$\bm{A} \bm{u}$. Therefore, any fast implementation for the evaluation of the discretized operator can be applied inside the smoother and parallel scalability is directly inherited from the operator evaluation. Algorithm~\ref{alg:Chebyshev} details the Chebyshev iteration with iteration index~$j$ and~$n_{\mathrm{s}}$ smoothing steps, where the two additional scalar parameters~$\sigma_j$ and~$\theta_j$ are calculated according to the theory of Chebyshev polynomials and require an estimation of the maximum eigenvalue~$\lambda_{\mathrm{max}}$ of~$\bm{A}$. The parameters are determined such that the Chebyshev smoother tackles eigenvalues in the range~$\left[ 0.06 \lambda_{\mathrm{max}}, 1.2\lambda_{\mathrm{max}}\right]$ on the current level, while smaller eigenvalues are damped by the coarse-grid correction. Since the maximum eigenvalue is only estimated, a safety factor of~$1.2$ is included to ensure robustness of the smoother. Note that the precise value used for the lower bound is not critical in terms of robustness and iteration counts. A Chebyshev iteration with~$n_{\mathrm{s}}$ pre- and post-smoothing steps is denoted as Chebyshev($n_{\mathrm{s}},n_{\mathrm{s}}$) in the following, typical values for which the smoother is most efficient being~$n_{\mathrm{s}} = 3, ..., 6$, see for example~\cite{Kronbichler2019}. As a default parameter, we use~$n_{\mathrm{s}} = 5$ and show additional results in form of a parameter study in Section~\ref{sec:Results}.

The diagonal required by the Chebyshev smoother is calculated in the setup phase. The maximum eigenvalue needed by the Chebyshev iteration is estimated by performing~$20$ conjugate gradient iterations. Compared to a single solution of the linear system, this cost is not negligible. However, for many large-scale time dependent problems where~$\mathcal{O}(10^5-10^7)$ time steps have to be solved, setup costs are amortized after a few time steps, which is why we do not explicitly consider these costs in the present work. For details on setup costs see, e.g.,~\cite{Kronbichler2019}. This is further justified by the fact that the costs are proportional to the costs of a fine-level matrix-vector product, and therefore increase similarly under mesh refinement as the solution of the linear system of equations itself. While the present work is restricted to the constant-coefficient Poisson case, it should be mentioned that Chebyshev smoothing has been reported to work well also for variable-coefficient problems with smoothly varying coefficient~\cite{Kronbichler2018, Sundar2015}.

\subsection{Coarsening strategies and multigrid transfer operations}\label{sec:Transfer}
\begin{figure}[!ht]
\centering
  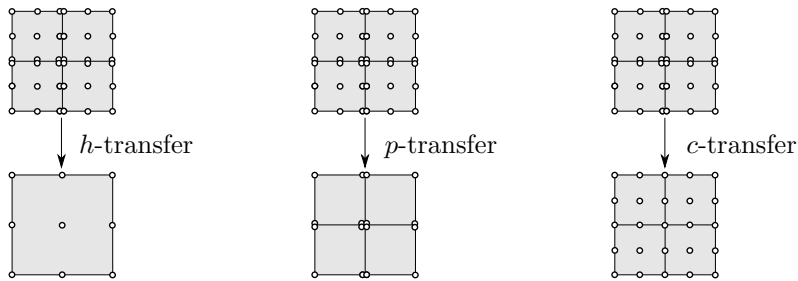
\caption{Illustration of elementary coarsening strategies for nodal high-order discontinuous Galerkin discretizations.}
\label{fig:CoarseningStrategies}
\end{figure}

The multigrid level~$l$ introduced in Algorithm~\ref{alg:VCycle} is uniquely defined by the grid size~$h$, the polynomial degree~$p$, and the continuity parameter~$c \in \left\{\mathrm{DG},\mathrm{FE}\right\}$
\begin{align*}
l = f(h, p, c) \; .
\end{align*}
From one multigrid level to the next, only one of the three parameters may change for the hybrid multigrid methods discussed in this work. For example, a transfer from DG space to FE space leads to two multigrid levels that coincide with respect to grid size~$h$ and polynomial degree~$p$, i.e., a combined coarsening from high-order discontinuous space to low-order continuous space is not considered here. The approach is denoted as~$h$-/$p$-multigrid if geometric/polynomial coarsening is employed only. Combined geometric-polynomial multigrid is denoted as~$hp$- or~$ph$-multigrid, depending on which coarsening is applied first,~\footnote{We only consider sequential coarsening in~$h$ and~$p$ as opposed to, e.g.,~\cite{Antonietti2015}, where the term~$hp$-multigrid is used for simultaneous coarsening in both~$h$ and~$p$ from one level to the next.} as illustrated in Figure~\ref{fig:HybridMultigrid}. Following this notation and depending on whether the DG-to-FE transfer is performed at high degree or at low degree, we denote this approach as~$cp$-multigrid or~$pc$-multigrid, respectively, or as~$ch$-multigrid or~$hc$-multigrid if geometric coarsening is involved. Applying all three possibilities for coarsening would for example result in a~$cph$-multigrid strategy, with the~$c$-coarsening performed first, followed by~$p$-coarsening and finally~$h$-coarsening. The different types of coarsening are illustrated in Figure~\ref{fig:CoarseningStrategies}. In all cases algebraic multigrid may be applied as a coarse-grid solver.

We write the prolongation of the coarse-level correction from coarse to fine levels generically (for all types of transfers~$t \in \left\{ h, p, c\right\}$) as
\begin{align}
\bm{u}^{(l)} = \bm{P}^{l}_{l-1} \bm{u}^{(l-1)} = \sum_{e=1}^{N_{\mathrm{el}}^{(l)}} \bm{S}_e^{l} \bm{P}_{e}^{l,l-1} \bm{G}_e^{l-1}  \bm{u}^{(l-1)} \ ,
\end{align}
where the global prolongation operator is expanded into the sum over all elements on the fine level with elementwise prolongation operator~$\bm{P}_{e}^{l,l-1}$. The gather operator~$\bm{G}_e^{l-1}$ extracts local degrees of freedom of a coarse-level element from the global DoF vector. The scatter operator~$\bm{S}_e^{l}$ coordinates the write of local degrees of freedom into the global fine-level DoF vector and additionally performs a weighting of degrees of freedom according to the multiplicity of a shared node in case of continuous function spaces. The elementwise prolongation operator is realized as~$L^2$-orthogonal projection
\begin{align}
\left(v_{h}^{(l)},u_{h}^{(l)}\right)_{\hat{\Omega}_{e}^{(l)}} = \left(v_{h}^{(l)},u_{h}^{(l-1)}\right)_{\hat{\Omega}_{e}^{(l)}} \rightarrow \bm{P}_{e}^{l,l-1} = \left(\bm{M}^{l}_e\right)^{-1} \bm{M}_e^{l,l-1} \ ,
\end{align}
where~$\bm{M}_e^l$ denotes the mass matrix and~$\bm{M}_e^{l,l-1}$ the embedding from space~$l-1$ into~$l$. Note that the integral is performed in reference space over the fine-level element~$\hat{\Omega}_{e}^{(l)}$. Therefore, the operation is the same for all elements and is done only once in the setup phase where the 1D prolongation matrices are constructed. Prolongation in multiple dimensions is constructed as the tensor product of 1D operations, exploiting fast matrix-free evaluation techniques. The 1D prolongation matrices represent the interpolation of coarse-level basis functions into the nodes of the fine-level basis functions. In the case of~$h$-coarsening and for general mappings from reference to physical space, however, the coarse-level space is no longer a subset of the fine-level space. Therefore, the chosen multigrid transfer operations implicitly introduce the approximation of nested function spaces as also mentioned, e.g., in~\cite{Lu2014}. In case of~$p$-transfer and~$c$-transfer, the function spaces are ``strictly'' nested. Restriction of the residual~$\bm{r}$ onto coarser levels is defined as the transpose of prolongation,
\begin{align}
\bm{r}^{(l-1)} = \bm{R}^{l-1}_{l} \bm{r}^{(l)} = \left(\bm{P}^{l}_{l-1}\right)^{\mathsf{T}} \bm{r}^{(l)} = \sum_{e=1}^{N_{\mathrm{el}}^{(l)}} \left(\bm{G}_{e}^{l-1}\right)^{\mathsf{T}} \left(\bm{P}_{e}^{l,l-1}\right)^{\mathsf{T}} \left(\bm{S}_{e}^{l}\right)^{\mathsf{T}} \bm{r}^{(l)} \ .
\end{align}

\subsubsection{$h$-coarsening}
A hierarchy of $h$-levels is constructed based on the octree concept, see for example~\cite{Sundar2012, Burstedde2011} for details on aspects of the chosen mesh topology. Therefore, coarser meshes in the multigrid context are not obtained by coarsening a fine mesh, but rather by starting from a coarse mesh that is refined uniformly several times to obtain the fine mesh. This coarse mesh also forms the coarse-grid problem in the multigrid algorithm. From this perspective, it is clear that pure~$h$-multigrid based on the octree approach works well for cube-like domains of moderate geometrical complexity, but reaches limitations for complex geometries where only one or two refinement levels applied to the coarse mesh might be affordable in practice. In these cases, it is essential to further coarsen the problem by the use of~$p$-multigrid and algebraic multigrid techniques described in more detail below. Here, we restrict ourselves to meshes without hanging nodes and each octree has the same number of mesh refinement levels. Multigrid methods for high-order discretizations on adaptively refined meshes are discussed in~\cite{Kanschat2004, Janssen2011, Kronbichler2018, Kronbichler2019, Sundar2012} in a pure~$h$-multigrid context.

\subsubsection{$p$-coarsening}

As opposed to~$h$-multigrid,~$p$-multigrid offers the possibility for various~$p$-coarsening strategies. Reducing the polynomial degree by one,~$p_{l-1} = p_l-1$, is frequently applied in literature~\cite{Bassi2003,Fidkowski2004,Fidkowski2005,Nastase2006JCP,Shahbazi2009,Diosady2009,Bassi2009,Bassi2011,Ghidoni2014, Antonietti2015, Fabien2019}. An alternative is to reduce the polynomial degree by a factor of two,~$p_{l-1} = p_l/2$ (with appropriate rounding operation), which has been used in~\cite{Helenbrook2003,Helenbrook2008,Mascarenhas2009,Mascarenhas2010,Helenbrook2016}. This coarsening strategy has a close analogy to~$h$-coarsening since the number of degrees of freedom is reduced by a factor of two in each coordinate direction from one level to the next. It is also not uncommon to immediately reduce the polynomial degree to its minimum,~$p_{l-1} = 0$ or~$p_{l-1} = 1$ for all~$p_l$ (two-level algorithm), see for example~\cite{Rasetarinera2001, Persson2008, Bastian2019}. Elementwise constant shape functions with $p_{l=0} = 0$ are not considered in this work. On the one hand, the present DG discretization is not consistent for polynomial degree~$p=0$. On the other hand, as argued in~\cite{Helenbrook2008}, the small-wave-number modes that remain after smoothing are essentially continuous for diffusive problems and are, therefore, not well represented by a piecewise constant coarse space with~$p=0$. For the neutron diffusion problems studied in~\cite{OMalley2017}, a continuous~$p=1$ coarse space has been found to be advantageous over a piecewise constant space. It has been observed in~\cite{Persson2008,Shahbazi2009} by the example of the compressible Navier--Stokes equations involving diffusive terms that~$p_{l=0}=1$ performs better than~$p_{l=0}=0$. A piecewise constant space with~$p_{l=0}=0$ is typically used in the convection-dominated limit and the compressible Euler equations~\cite{Luo2006,Nastase2006JCP}. In~\cite{Bastian2019},~$p_{l=0}=0$ is also used for a Poisson problem with variable coefficients. These previous studies indicate that the optimal coarse space depends on the model problem under investigation. Since the present work is restricted to the constant-coefficient Poisson problem, we also restrict ourselves to a specific choice~$p_{l=0}=1$ for the coarse space. Discussions and comparisons of different $p$-sequences can be found in~\cite{Helenbrook2008} in the context of iteration counts and in~\cite{Nastase2006JCP} in terms of iteration counts and computational costs. In that work, only a single polynomial degree of~$p=4$ is investigated. Here, we foster a more rigorous investigation of the following $p$-coarsening strategies
\begin{itemize}
\item $p_{l-1} = p_l-1$ (decrease by one),
\item $p_{l-1} = \lfloor p_l/2 \rfloor $ (bisection),
\item $p_{l-1} = 1 \, \forall p_l$ (two-level algorithm),
\end{itemize} 
considering a wide range of polynomial degrees~$p$ and studying the impact on both iteration counts and computational costs. All~$p$-levels are exploited in our multigrid algorithm until~$p=1$ is reached.

\subsubsection{$c$-coarsening (transfer from discontinuous to continuous space)}

A transfer from the discontinuous space to a continuous space at the coarse degree~$p=1$ is used in~\cite{Helenbrook2008, OMalley2017}, an idea that has already been described in~\cite{Lasser2001} in the context of two-level overlapping preconditioners. A transfer at the highest degree~$p$ is suggested in~\cite{Rudi2015} without justification and with results shown only for the lowest degree~$p=1$. This approach might be counter-intuitive at first sight since an additional multigrid level at high polynomial degree (and therefore with expensive smoothers) is introduced and the problem size is only marginally reduced for a DG-to-FE transfer at high degree, i.e., by a factor of~$\left(1 + 1/{p}\right)^d$. It is interesting to note that a similar idea called conforming aggregation is used in~\cite{Olson2011} in the context of smoothed aggregation algebraic multigrid techniques where degrees of freedom at the same spatial location are aggregated on the finest level. For the two-level scheme proposed in~\cite{Dobrev2006, Bastian2012, Siefert2014, Bastian2019}, the high-order DG space is directly reduced to a linear conforming space. According to our experiments, this could be the reason for the strong increase in iteration counts observed in~\cite{Siefert2014, Bastian2019} for increasing~$p$ (and for a similar two-level preconditioner used in~\cite{Remacle2016}). As mentioned previously, we introduce an additional multigrid level for the DG-to-FE transfer as in~\cite{OMalley2017}, i.e., the transfer to continuous FE space is done at constant degree~$p$ and mesh size~$h$ and we found that this is important for optimal multigrid convergence rates. We investigate two variants of the DG-to-FE transfer in this work, namely performing this transfer at highest degree or lowest degree~$p=1$ (and similarly on the finest mesh or coarsest mesh). Performing the transfer to continuous elements on the finest level has very attractive properties. It reduces the iteration counts considerably, and yields a multigrid solver for SIPG discretizations of the Poisson equation that is robust w.r.t.~the penalty parameter of the SIPG method. Theoretical background for this behavior is provided in~\cite{Antonietti2017}, where this approach is motivated from the perspective of space splitting and auxiliary space methods. The important difference is that we here integrate this spliting into multigrid with the same smoother used on all levels.

The elementwise prolongation matrix is an identity matrix in the case of a DG-to-FE transfer since the continuous and discontinuous function spaces are the same from an elementwise perspective. Accordingly, the degrees of freedom shared by neighboring elements in the continuous case are simply injected into the degrees of freedom duplicated in the discontinuous case. With restriction being the transposed operation, the residual contributions of degrees of freedom of duplicated nodes in the discontinuous case are summed into the uniquely defined degree of freedom in the continuous case.

\begin{remark}
The two-level approaches in~\cite{Lasser2001, Dobrev2006, Antonietti2017} are also known or interpreted as auxiliary space preconditioning. We refrain from this nomenclature in the present work and rather categorize these approaches as one type of multigrid coarsening in the generalized framework of hybrid multigrid algorithms. The multigrid methods in~\cite{Bastian2012, Siefert2014} are introduced as algebraic multigrid methods that are ``not fully algebraic''. In the present work, we foster a fine-level point of view and categorize these approaches as~$p$-multigrid (potentially with additional~$c$-coarsening) with algebraic multigrid applied as coarse-grid solver; for good reasons, because the fine levels are those where the numerical method spends its time (assuming that the method is applied away from the strong scaling limit) and are those that determine the computational efficiency of the approach.
\end{remark}

\subsection{Coarse-grid solver}
The success of multigrid methods originates from the fact that the coarse-grid correction ensures mesh-independent convergence rates as well as low iteration counts and -- at the same time -- causes only low computational overhead as compared to the operations on the fine level. It is therefore important that the coarse-grid correction does not deteriorate the multigrid convergence rate which should only be affected by the smoother on the fine level. This is particularly important for the AMG coarse-grid solver that does not necessarily converge at the same rate as the smoothers on the geometric levels of the multigrid hierarchy would allow to. For this reason, it is reasonable to solve the coarse-level problem by an inner Krylov solver preconditioned by an AMG V-cycle to a specified tolerance instead of only a single AMG V-cycle as coarse-grid solver. Note that using a Krylov solver within the preconditioner does no longer guarantee that the preconditioner is a stationary operation, which might require the use of flexible solvers in general. Since we did not observe convergence problems in the present work, a basic CG iteration is used throughout as outer Krylov solver. Extending AMG solvers designed for continuous discretizations to the discontinuous case is not trivial without further measures as shown in~\cite{Olson2011}. Since we want to apply the AMG coarse-grid solver in a black-box fashion in its optimal regime, we mainly show performance numbers for AMG applied to a continuous discretization of the coarse problem with lowest degree~$p=1$, see also~\cite{Bastian2012, Siefert2014}. 
The present work makes use of the AMG implementation provided by the~\texttt{Trilinos ML} project~\cite{TrilionsML2006}, using one V-cycle with one smoothing step of an~\texttt{ILU} smoother without fill-in and no overlap (i.e., in a block-Jacobi fashion over the MPI ranks) and an~\texttt{Amesos-KLU} coarse solver unless specified otherwise. A comparative study of different AMG solver frameworks is beyond the scope of this study, and is for example shown in~\cite{OMalley2017} for the neutron diffusion equation, or in~\cite{Offermans2019} in the context of computational fluid dynamics.

\section{Matrix-free operator evaluation}\label{sec:MatrixFree}
The overall performance of the multigrid solver crucially depends on the speed at which the matrix-vector product~$\bm{A}\bm{u}$ can be performed. The outer Krylov solver, the multigrid V-cycle, and also the multigrid smoothers only require the action of the linear operator~$\bm{A}$ applied to a vector. Since multigrid transfer operators can also be realized in a matrix-free way using sum-factorization, all components of the algorithm outlined in Section~\ref{sec:HybridMultigrid} (apart from the AMG coarse-grid solver) are amenable to fast matrix-free operator evaluation. The present work builds on matrix-free evaluation routines using the implementation developed in~\cite{Kronbichler2012, Kronbichler2019fast} and available in the~\texttt{deal.II} finite element library~\cite{dealII90}. The global matrix-vector product is written as a loop over all elements and faces with the local weak form evaluated by numerical quadrature 

\begin{align}
\bm{A}\bm{u} = \sum_{e=1}^{N_{\mathrm{el}}} \bm{S}_e \bm{I}_e^\mathsf{T} \bm{D}_e \bm{I}_e \bm{G}_e \bm{u} +  \sum_{f=1}^{N_{\mathrm{faces}}} \bm{S}_f \bm{I}_f^\mathsf{T} \bm{D}_f \bm{I}_f \bm{G}_f \bm{u} \; . \label{eq:MatrixFreeAbstractNotation}
\end{align}
For volume integrals over~$\Omega_e$, the gather operation~$\bm{G}_e$ extracts the local degrees of freedom associated to element~$e$,~$\bm{u}_e = \bm{G}_e \bm{u}$. Similarly, for the integral over a face~$f = \partial \Omega_{e^-} \cap \partial \Omega_{e^+}$, ~$ \bm{G}_f$ extracts the relevant degrees of freedom of the two elements~$e^-, e^+$ required for the computation of the face integral,~$ \left(\bm{u}_{e^-}^\mathsf{T}, \bm{u}_{e^+}^\mathsf{T}\right)^\mathsf{T} = \bm{G}_f \bm{u}$. Then, the computation of volume and face integrals is a 3-step process described by~$\bm{I}_e^\mathsf{T} \bm{D}_e \bm{I}_e$ and~$\bm{I}_f^\mathsf{T} \bm{D}_f \bm{I}_f$, respectively. This 3-step process forms the core of the matrix-free operator evaluation and is explained in more detail below. Finally, the scatter operations~$\bm{S}_e = \bm{G}_e^\mathsf{T}$ and~$\bm{S}_f = \bm{G}_f^\mathsf{T}$ add contributions of volume and face integrals into the global residual vector according to the mapping of local-to-global degrees of freedom. Independently of the specific discretization technique, matrix-free techniques are the state-of-the-art implementation for high-performance realizations of PDE solvers, see for example~\cite{Rudi2015, Ichimura2015, Gmeiner2015}.

\begin{figure}[!ht]
\centering
  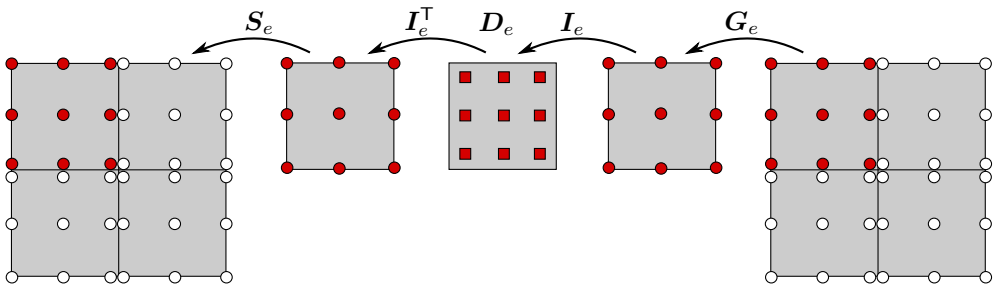
\caption{Illustration of matrix-free operator evaluation for the computation of cell integrals for a discontinuous, nodal basis with degree~$p=2$ and~$p+1=3$ interpolation and quadrature points per coordinate direction. Note that this illustration shows the non-vectorized case with the volume integral performed for a single element only.}
\label{fig:IllustrationMatrixFree}
\end{figure}

Next, we detail the procedure for the matrix-free operator evaluation for the volume integral over~$\Omega_e$
\begin{align*}
(\bm{A}_e \bm{u}_e)_i &= \intele{\Grad{N_i}}{\Grad{u_h}} = \int_{\Omega_e} (\nabla_{\bm{x}} N_i)^\mathsf{T} \nabla_{\bm{x}} u_h^e d{\bm{x}}\\
& = \int_{[0,1]^d} ({\bm{J}^e}^{-\mathsf T}\nabla_{\boldsymbol{\xi}} N_i)^\mathsf{T} ({\bm{J}^e}^{-\mathsf T}\nabla_{\boldsymbol{\xi}} u_h^e)\vert \det \bm{J}^e \vert d\boldsymbol{\xi}\\
&\approx
\sum_q \underbrace{(\nabla_{\boldsymbol{\xi}} N_i(\xi_q))^{\mathsf{T}}}_{\left(\bm{I}_e^\mathsf{T} \right)_{i,q}}
\underbrace{{\bm{J}_q^e}^{-1} (w_q \vert\det \bm{J}_q^e\vert)  {\bm{J}_q^e}^{-\mathsf T}}_{\left(\bm{D}_e\right)_{q,q}} 
\sum_j \underbrace{\nabla_{\boldsymbol{\xi}} N_j(\xi_q)}_{\left(\bm{I}_e\right)_{q,j}} u_j^e \\
 & = \left(\bm{I}_e^\mathsf{T} \bm{D}_e \bm{I}_e \bm{u}_e\right)_i, \forall i=1,\ldots, (p+1)^d \; ,
\end{align*}
The integral over the physical domain is first transformed to the reference element, giving rise to geometry terms such as the Jacobian~${\bm{J}^e}$. Integration is then performed by Gaussian quadrature, introducing the quadrature weight~$w_q$ and replacing the integral by a sum over all quadrature points. The last row shows how the elementwise computation of integrals can be interpreted in terms of the more abstract notation introduced in equation~\eqref{eq:MatrixFreeAbstractNotation}. The interpolation operator~$\bm{I}_e$ computes the gradient (in reference coordinates) of the solution at all quadrature points by interpolation of the basis functions according to the polynomial expansion introduced in Section~\ref{sec:ModelProblem}. The differential operator~$\bm{D}_e$ applies the PDE operator for all quadrature points and depends on data associated to the current element~$e$ for non-Cartesian element geometries. The integration operator~$\bm{I}_e^\mathsf{T}$ multiplies by the gradient of the test function and sums over all quadrature points~(=integration). It can be easily seen from the above equation that the integration step is the transpose of the interpolation step. Interpolation and integration are done in reference coordinates and do not depend on the current element~$e$. To obtain optimal computational complexity, it is essential to exploit the tensor product structure of the shape functions in the interpolation and integration steps. This optimization technique is called sum-factorization and replaces the sum over all nodes~$j$ by~$d$ sums over the one-dimensional nodes~$j_1, \ldots, j_d$, leading to a complexity of~$(p+1)^{d+1}$ operations. Applying~$d$ one-dimensional interpolation kernels for~$d$ gradients gives rise to~$d^2$ kernels. However, the operations can be reduced to~$2d$ kernels by first interpolating into a collocation basis~($d$ kernels) and subsequently evaluating the gradient in the collocation basis (another~$d$ kernels)~\cite{Kronbichler2019fast}. Another optimization technique reducing the number of operations for the one-dimensional kernels exploits the symmetry of the one-dimensional shape functions and is called even-odd decomposition~\cite{Kopriva2009}. An illustration of the matrix-free evaluation process is provided in Figure~\ref{fig:IllustrationMatrixFree}. The computation of face integrals follows the same principles and we refer the interested reader to~\cite{Kronbichler2012, Kronbichler2019fast} for more details. 
For the special case of affine element geometries, a single Jacobian~$\bm{J}^e$ is used at all quadrature points of a cell. For deformed elements, a separate Jacobian~$\bm{J}^e_q$ is precomputed for each quadrature and stored as~$\left(\bm{J}^e_q\right)^{-\mathsf T}$, which is then accessed during the operator evaluation and represents the main memory traffic. On faces, the quantity~$\boldsymbol n^\mathsf T \left(\bm{J}^e_q\right)^{-\mathsf T}$ is pre-computed at each quadrature points. We refer to~\cite{Kronbichler2019fast} for possible alternatives regarding the evaluation of the geometry. Apart from the operator evaluation and smoothing on all multigrid levels, also the multigrid transfer operators discussed in Section~\ref{sec:Transfer} are implemented with optimal-complexity matrix-free algorithms.

\paragraph{Vectorization over elements and faces} The matrix-free operator evaluation performs the same operations for all elements, the only difference is that integrals over different elements operate on different parts of the solution vector~$\bm{u}$ and the geometry information~$J_q^e$ has to be stored and loaded separately for each element in case of deformed element geometries. In order to exploit the single-instruction-multiple-data (SIMD) vectorization capabilities of modern hardware with wide SIMD units, the present implementation groups together several elements or faces and performs the integrals in the weak form concurrently for this batch of elements or faces. This technique has first been proposed in~\cite{Kronbichler2012}. The basic data type for the operations in the matrix-free evaluation process is therefore \texttt{VectorizedArray<Number>}, with~\texttt{Number} being a template for a~\texttt{C++} data type such as~\texttt{double} or~\texttt{float}. For the hardware used in the present work with support for AVX512 (see Table~\ref{tab:Hardware}), vectorization is done over~$8$ elements/faces in double precision and~$16$ in single precision. For meshes with the number of elements/faces not being a multiple of the vectorization width, parts of the vectorized array remain empty for these corner cases.

\paragraph{Complexity and throughput} The theoretical complexity of the matrix-free evaluation is~$\mathcal{O}((p+1)^{d+1})$ operations and~$\mathcal{O}((p+1)^{d})$ data, resulting in a linear complexity,~$\mathcal{O}(p^{1})$, or constant complexity,~$\mathcal{O}(1)$, per degree of freedom, depending on whether arithmetics or memory transfer forms the main bottleneck. On modern hardware with high Flop-to-Byte ratios, the present matrix-free implementation tends to be memory-bound when implemented with a minimum of arithmetic operations~\cite{Kronbichler2019fast}. Figure~\ref{fig:throuhgput_matrix_free_vs_matrix_based} shows the throughput of the present implementation measured for the evaluation of the scalar Laplace operator on a 3D cube geometry with periodic boundary conditions for both Cartesian and curved elements. In practice, the throughput measured in degrees of freedom per second depends only mildly on the polyomial degree and suggests an almost constant complexity up to moderately high polynomial degrees. The fact that the observed complexity is significantly better than the theoretical complexity of volume integrals can be explained by the fact that face integrals as well as data access (with constant complexity per unknown) are performance relevant for moderately high polynomial degrees.

\paragraph{Mixed-precision multigrid} The matrix-free algorithm outlined above is perfectly suited for mixed-precision computations in the multigrid preconditioned Krylov solver, following the idea of~\cite{Gropp2000}. This is due to the fact that the amount of data transferred from main memory reduces by a factor of two in case of single precision (implying twice the throughput in terms of elements processed per time), and the vectorization strategy with explicit vectorization over elements/faces also allows twice the throughput in terms of arithmetics. The throughput of the matrix-vector product shown in Figure~\ref{fig:throuhgput_matrix_free_vs_matrix_based} is therefore raised by a factor of approximately~$2$ when reducing accuracy from double precision to single precision. To not spoil accuracy of the numerical approximation of the solution and ensure convergence of the outer Krylov solver, single precision is only used in the multigrid V-cycle. The outer Krylov solver operates in double precision. Larger round-off errors in the multigrid cycle can be tolerated since these high-frequency errors introduced by single-precision round-off errors are tackled by the multigrid smoothers~\cite{Kronbichler2019} and since multigrid is only a preconditioner applied to the residual of the outer Krylov solver, see Algorithm~\ref{alg:CG}. Since the~\texttt{Trilinos ML} solver used here only supports double precision, the AMG coarse-grid preconditioner operates in double precision. The performance of mixed-precision is compared to pure double-precision computations in Figure~\ref{fig:throuhgput_cube_1core_vs_1node} below and discussed in Section~\ref{sec:Results}.

\section{Results}\label{sec:Results}
We introduce relevant performance metrics used to evaluate the efficiency of the present hybrid multigrid methods in Section~\ref{sec:Metrics}. Information on the hardware under consideration is given in Section~\ref{sec:Hardware}. The considered test cases are briefly summarized in Section~\ref{sec:TestCases}, before numerical results are shown for each problem in the subsequent sections.

\subsection{Performance metrics}\label{sec:Metrics}

Frequently used metrics are the average multigrid convergence rate~$\rho$ and the number of iterations~$n_{10}$ needed to reduce the residual by ten orders of magnitude ($\varepsilon_{10} = \Vert \bm{r}_{n_{10}} \Vert_2 / \Vert \bm{r}_0 \Vert_2 = 10^{-10}$)
\begin{align*}
\rho = \left(\frac{\Vert \bm{r}_n \Vert_2}{\Vert \bm{r}_0 \Vert_2} \right)^{\frac{1}{n}} \; , \; n_{10} = \frac{\log_{10} \left(\Vert \bm{r}_{n_{10}} \Vert_2 / \Vert \bm{r}_0 \Vert_2 \right)}{\log_{10} \rho} = \frac{-10}{\log_{10} \rho} \; ,
\end{align*}
where~$ \bm{r}_n$ denotes the residual after~$n$ iterations. These quantities are well suited to demonstrate mesh-independent convergence rates, or to quantitatively investigate robustness of the multigrid method, i.e., the influence of certain parameters such as mesh anisotropies, variable coefficients, or the polynomial degree on the convergence behavior of the multigrid algorithm. However, they are not suited to quantify the effectiveness of smoothers in terms of computational efficiency. 
To measure computational costs, theoretical measures such as operation counts required for the matrix-vector product or matrix nonzeros are often considered~\cite{Lottes2005,Sundar2015}. These quantities should be considered with some care because they inherently contain assumptions on the bottleneck (for example that the algorithm is compute-bound so that floating point operations are really a cost measure). However, this depends on many aspects such as the hardware under consideration (Flop-to-Byte ratio), the implementation strategy (matrix-based vs. matrix-free), and the optimization level of the implementation. For example, it is important to implement the matrix-free algorithms discussed here with a minimum of operations and to exploit SIMD capabilities of modern hardware. Due to these uncertainties and model assumptions of theoretical cost measures, we prefer experimental cost measures determined from the actual performance of the multigrid solver, in the spirit of~\cite{Kronbichler2018,Bastian2019}. An effective number of fine-level matrix-vector products, denoted as~$n_{10, \mathrm{mat-vec}}$ in the following, is helpful to incorporate computational costs for the smoother and to compare different smoothers in the metric of computational costs instead of global iteration counts. It is unclear whether more aggressive matrix-based smoothers resulting in lower iteration counts are also superior in the practically relevant time-to-solution metric. The quantity~$n_{10, \mathrm{mat-vec}}=t_{\mathrm{wall}, \bm{u}= \bm{A}^{-1}\bm{b} (\varepsilon_{10})}/t_{\mathrm{wall}, \mathrm{mat-vec}}$ is defined as the ratio of the wall time for one application of the linear solver with tolerance~$\varepsilon_{10}$ over the wall time for one operator evaluation. Since absolute wall times depend on the problem size, it is useful to express~$n_{10, \mathrm{mat-vec}}$ as a function of two normalized quantities. The first one is the efficiency~$E_{\mathrm{mat-vec}}$ of the matrix-free operator evaluation measured as the number of degrees of freedom~$N$ processed per second per core (also denoted as throughput)
\begin{align}
E_{\mathrm{mat-vec}} = \frac{N}{t_{\mathrm{wall}, \mathrm{mat-vec}} N_{\mathrm{cores}}} \, .
\end{align}
The second one is the time~$t_{10}$ required by the multigrid solver to solve one degree of freedom per core with a residual reduction of~$\varepsilon_{10}$
\begin{align}
t_{10} = \frac{t_{\mathrm{wall}, \bm{u}= \bm{A}^{-1}\bm{b} (\varepsilon_{10})} N_{\mathrm{cores}}}{N} \, ,
\end{align}
or equivalently the throughput~$E_{10}=1/t_{10}$ of the linear solver in degrees of freedom solved per second per core. Then, the effective number of fine-level matix-vector products is determined experimentally as
\begin{align}
n_{10, \mathrm{mat-vec}} = t_{10} E_{\mathrm{mat-vec}} = \frac{E_{\mathrm{mat-vec}}}{E_{10}} \ .
\end{align}
The aim of~$n_{10, \mathrm{mat-vec}}$ is to obtain a measure for the algorithmic complexity of the whole multigrid solver, but as independent of hardware and absolute performance numbers as possible. The definition of~$n_{10, \mathrm{mat-vec}}$ has similarities with the parallel textbook efficiency factor defined in~\cite{Gmeiner2015b}, with the important difference that we define one fine-level matrix-vector product as work unit instead of one fine-level smoothing step. Since the overall goal is optimizing time-to-solution and since the operator evaluation~$\bm{A}\bm{u}$ is the only algorithmic component re-occurring for practically all iterative solvers and preconditioners, it is important to use~$\bm{A}\bm{u}$ as work unit so that the algorithmic complexity of different smoothers is reflected in the values achieved for~$n_{10, \mathrm{mat-vec}}$. Furthermore, the performance advantage achieved by the use of mixed-precision multigrid is naturally included in our definition of~$n_{10, \mathrm{mat-vec}}$. Table~\ref{tab:Metrics} summarizes our performance metrics.

\begin{table}[t]
\caption{Performance metrics used to evaluate the computational efficiency of multigrid solvers.}
\label{tab:Metrics}
\renewcommand{\arraystretch}{1.1}
\begin{center}
\begin{tabular}{ll}
\hline
Quantity & description\\
\hline
$n_{10}$ & number of iterations to reduce the residual by ten orders of magnitude ($\varepsilon_{10}=10^{-10}$)\\
$t_{10}$ & wall time in seconds to solve one unknown per core to reach~$\varepsilon_{10}=10^{-10}$\\
$E_{10}$ & throughput of solver in unknowns solved per second per core ($= 1/t_{10}$)\\
$E_{\mathrm{mat-vec}}$ & throughput of matrix-free operator evaluation in unknowns processed per second per core\\
$n_{10, \mathrm{mat-vec}}$ & effective number of fine-level mat-vec products ($ = E_{\mathrm{mat-vec}} / E_{10}$) to reach~$\varepsilon_{10}$\\
\hline
\end{tabular}
\end{center}
\renewcommand{\arraystretch}{1}
\end{table}

\subsection{Hardware}\label{sec:Hardware}
The numerical experiments shown in this work are performed on an Intel Skylake architecture with AVX512 vectorization. Table~\ref{tab:Hardware} lists the specifications of the SuperMUC-NG supercomputer in Garching, Germany. The GNU compiler \texttt{g++} version 7.3 with optimization flags \texttt{-O3 -funroll-loops -std=c++17 -march=skylake-avx512} is used. All computations are done on thin nodes unless specified otherwise. The present analysis focuses mainly on the node-level performance because multigrid solvers are well-known to be scalable even to the largest supercomputers~\cite{Kronbichler2018,Gholami2016}. The multigrid communication is between nearest neighbors, both horizontally within the matrix-vector products and vertically between the multigrid levels with one round-trip per V-cycle through the coarse solver, assuming the latter is sufficiently cheap. This is backed up by performance projections to exascale machines where multigrid is expected to be primarily memory-limited within the nodes~\cite{Ibeid2018}. Good parallel scalability up to high core counts on large supercomputers when using AMG coarse-grid solvers is shown in~\cite{Offermans2019,Offermans2016}.

\begin{table}[t]
\caption{Performance specifications for hardware system of SuperMUC-NG at LRZ in Garching, Germany.}
\label{tab:Hardware}
\renewcommand{\arraystretch}{1.1}
\begin{small}
\begin{center}
\begin{tabular}{llll}
\hline
\multicolumn{2}{l}{Processor} & \multicolumn{2}{l}{Memory and Caches}\\
\hline
Processor type &  Intel Skylake Xeon Platinum 8174 & Memory per node (thin/fat) & 96/768 GByte\\
Frequency & 2.7 GHz & Theoretical memory bandwidth & 256 GByte/s\\
Cores per node & 48 & STREAM memory bandwidth & 205 GByte/s\\
SIMD width & 512 bit (AVX512) & Cache size (L2 + L3) per node & $2\cdot 57$  MByte \\
\hline
\end{tabular}
\end{center}
\end{small}
\renewcommand{\arraystretch}{1}
\end{table}

\subsection{Test cases}\label{sec:TestCases}
\begin{figure}[!ht]
\centering
\subfigure[Cube: Cartesian mesh (left) and section of curvilinear mesh (right) with~$8^3$ elements ($h/L=1/8$) and aspect ratios of~$\mathrm{AR}=1.0$ and~$2.9$, respectively.]{
\hspace{2cm}
\includegraphics[width=0.25\textwidth]{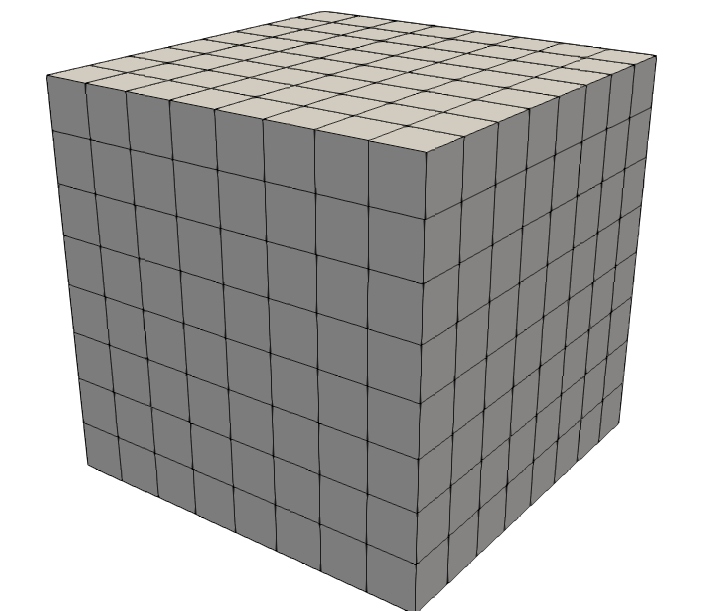}
\hspace{1cm}
\includegraphics[width=0.25\textwidth]{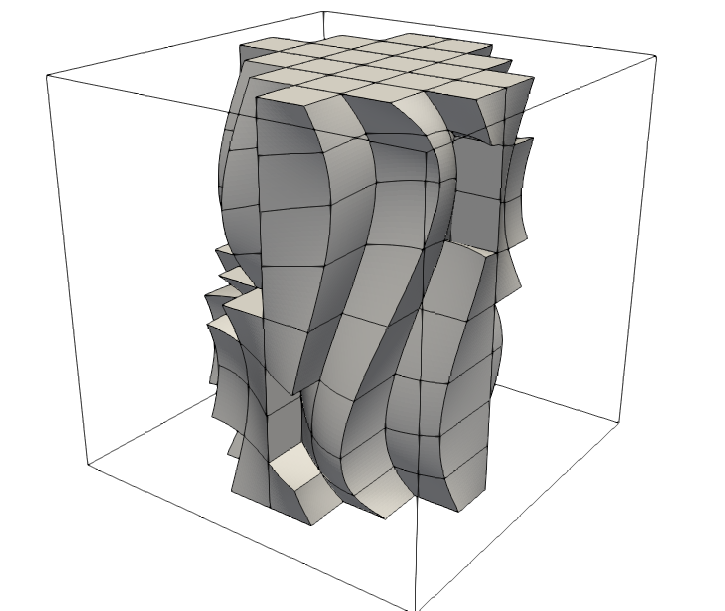}
\hspace{2cm}}
\subfigure[Nozzle: coarse mesh~$h_0$ of FDA nozzle geometry consisting of~$440$ elements~($\mathrm{AR} \approx 9.2$).]{
\includegraphics[width=0.75\textwidth]{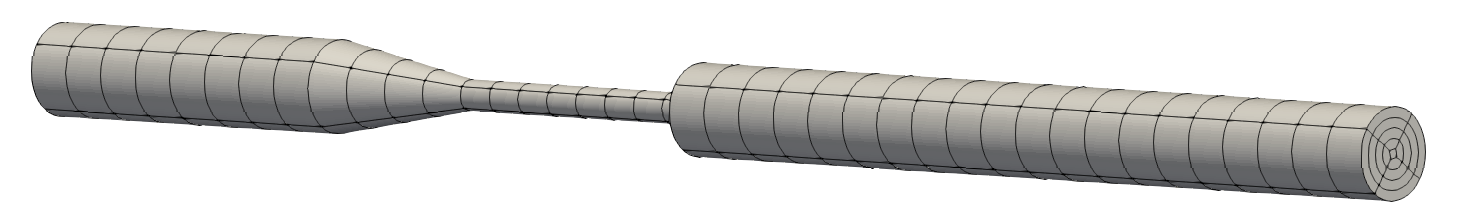}}
\subfigure[Lung: coarse mesh~$h_0$ of a patient-specific geometry of the human lung of a preterm infant for 6, 7, and 8 airway generations~(from left to right) with~$1968$,~$4236$, and~$9396$ elements, where the mesh with 8 generations has an aspect ratio of approximately~$\mathrm{AR} = 67$.]{
\includegraphics[width=0.25\textwidth]{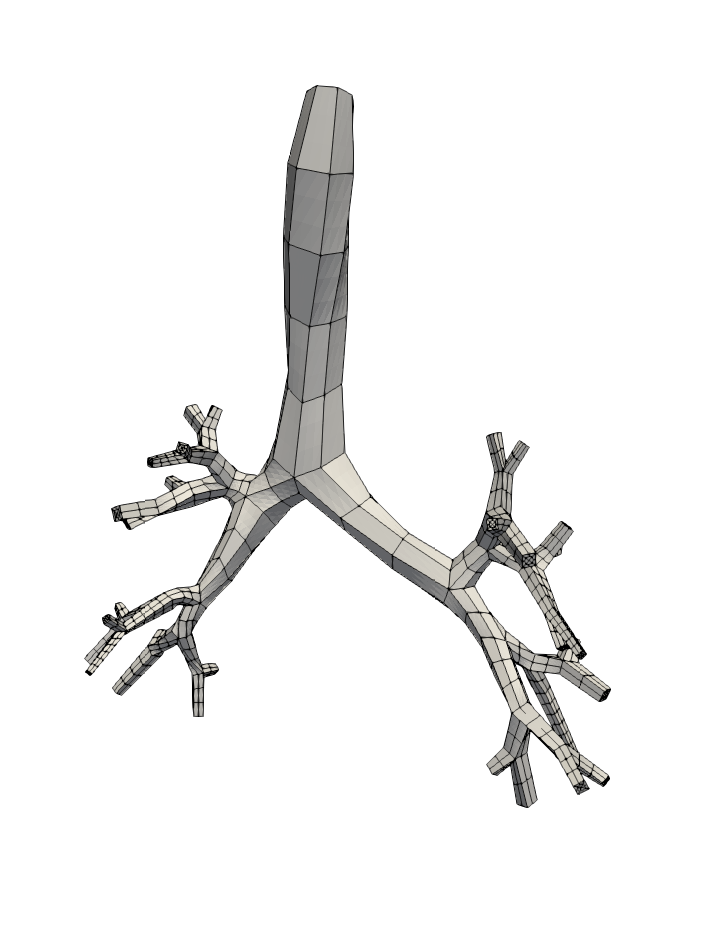}
\includegraphics[width=0.25\textwidth]{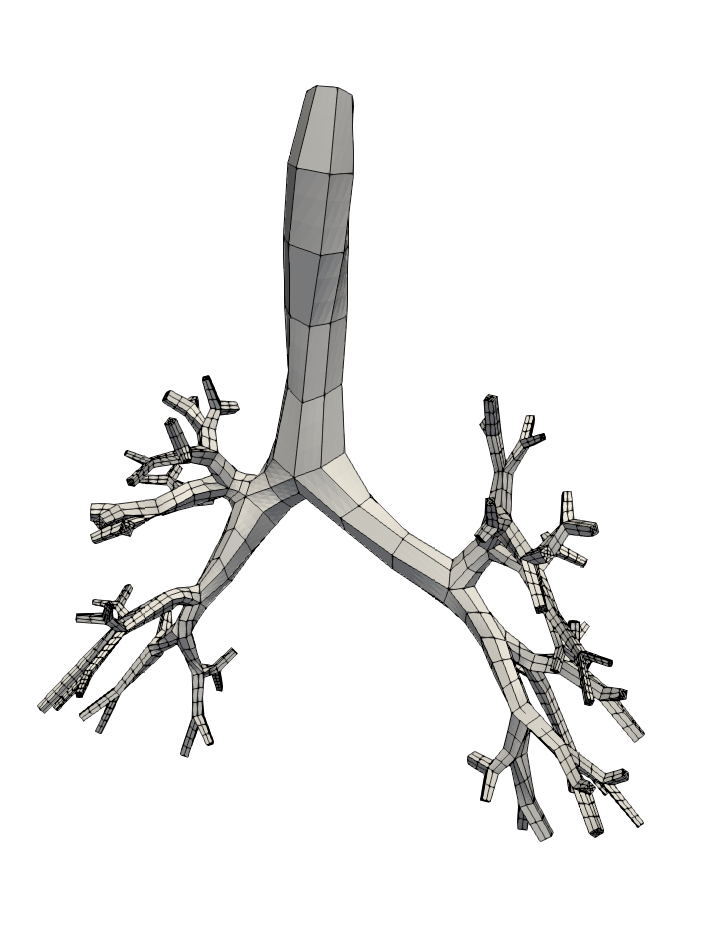}
\includegraphics[width=0.25\textwidth]{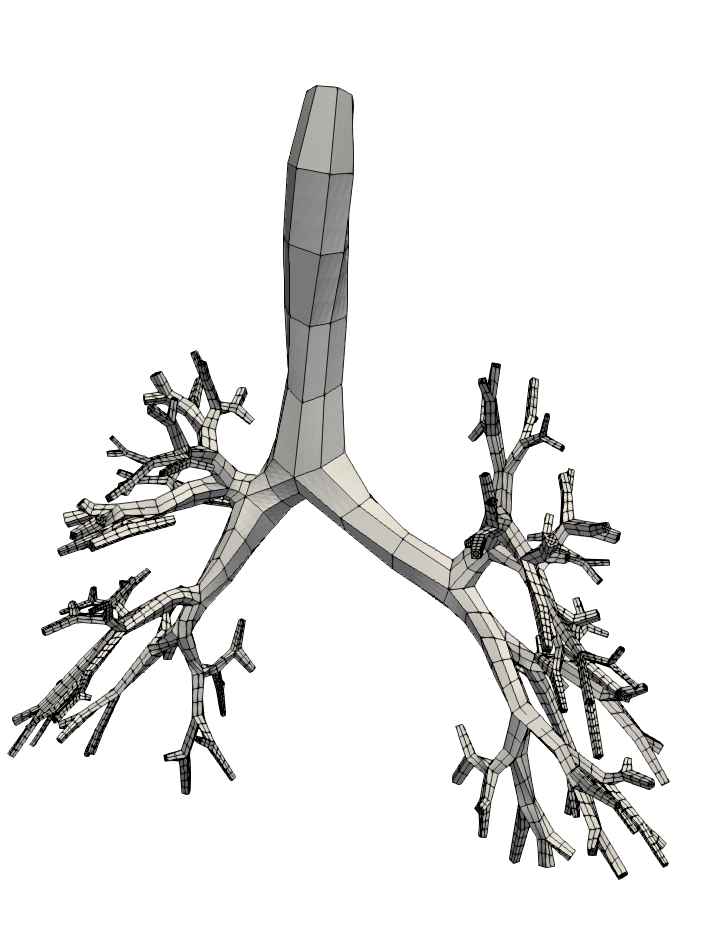}}
\caption{Visualization of geometries and meshes investigated in the present work. The size of the coarse-grid problem ranges from~$\mathcal{O}(10^1)$ to~$\mathcal{O}(10^4)$ elements.}
\label{fig:geometries_and_meshes}
\end{figure}

The proposed hybrid multigrid methods are investigated for a series of test cases with increasing complexity regarding the geometry and the number of elements on the coarse grid, as well as the maximum aspect ratio defined as
\begin{align}
\mathrm{AR} = \max_{e=1, \hdots, N_{\mathrm{el}}} \left(\max_{\Omega_e} \frac{\sigma_1(\bm{J}^e)}{\sigma_d(\bm{J}^e)}\right) \; ,
\end{align}
where~$\sigma_1$ and~$\sigma_d$ are the largest and smallest singular values of the Jacobian matrix~$\bm{J} = \partial \bm{x} /\partial \bm{\xi}$ (evaluated at all quadrature points of the element). A visualization of the geometries and the meshes of the different test cases is shown in Figure~\ref{fig:geometries_and_meshes}. We consider the following problems:
\begin{itemize}
\item \textit{Cube}: the geometry is a unit cube with~$\mathcal{O}(10^1)-\mathcal{O}(10^2)$ elements on the coarse grid. This geometry could also be discretized with a single element on the coarse grid, but we consider configurations with~$2^d, 3^d, 5^d$ elements on the coarse grid to test all multigrid components and make sure that the coarse-grid problem is non-trivial (but very small). This test case is well-suited to test the different multigrid ingredients, identify optimal multigrid coarsening strategies, perform parameter studies, study the impact of Cartesian and curved elements on iteration counts and throughput, and to compare the present implementation to the state-of-the-art (since data is mainly available for simple cube-like geometries in the literature).
\item \textit{Nozzle}: the geometry of this test case is the nozzle geometry of the FDA benchmark, which has been designed to assess CFD solvers for the simulation of the flow through medical devices~\cite{Malinauskas2017}. The geometry is a tube with gradual or sudden contractions/expansions of the cross section area, inducing separating flows and involving laminar, transitional, and turbulent flow regimes. We use a coarse-grid mesh with~$\mathcal{O}(10^3)$ elements in the present work. The tube and cone geometries are known analytically and used for high-order representations of the geometry via manifolds (using a cubic mapping). The blood flow through this device can be modeled as an incompressible flow, and the present work investigates the pressure Poisson component of the related incompressible Navier--Stokes solver. The mesh contains high-aspect-ratio elements with a moderate distortion, especially in the outflow part of the domain on the right.
\item \textit{Lung}: The most complex test case studied in this work is the geometry of the upper airway generations of the human lung, using a patient-specific geometry of a preterm infant, for which gas exchange mechanisms have been investigated recently in the literature~\cite{Roth2018}. The geometry is discretized with a purely hexahedral mesh and the coarse-grid problem consists of~$\mathcal{O}(10^4)$ elements for 8 airway generations. Simulating the flow of air through the human lung as a numerical solution of the incompressible Navier--Stokes equations again involves the solution of a pressure Poisson equation, which is studied in this work.
\end{itemize}

\begin{table}[!ht]
\caption{Summary of possible multigrid coarsening strategies. Regarding the nomenclature, the letters in the abbreviations are ordered according to the order in which the multigrid coarsening is performed from the fine level to the coarse level.}
\label{tab:SummaryCoarseningStrategies}
\renewcommand{\arraystretch}{1.1}
\begin{scriptsize}
\begin{center}
\begin{tabular}{ll}
\cline{1-2}
$h$-like MG & $p$-like MG\\
\hline
$h$   & $p$\\
$hp$  & $ph$\\
$hpc$ & $phc$\\
$hc$  & $pc$\\
$hcp$  & $pch$\\
$ch$  & $cp$\\
$chp$ & $cph$\\
\hline
\end{tabular}

\end{center}
\end{scriptsize}
\renewcommand{\arraystretch}{1}
\end{table}

\subsection{Cube}
We consider a simple analytical test case with solution
\begin{align*}
u(\bm{x}) = \sin (3\pi x_1) \sin (3\pi x_2) (3\pi x_3)
\end{align*}
on a cube geometry in 3D,~$\Omega = [-1, 1]^3$. Dirichlet boundary conditions are prescribed on the domain boundary using the known analytical solution. The right-hand side is determined according to the method of manufactured solutions,~$f(\bm{x}) = - \nabla^2 u(\bm{x}) = 27 \pi^2 \sin (3\pi x_1) \sin (3\pi x_2) (3\pi x_3)$. We analyze both a Cartesian mesh and a curvilinear mesh with deformation
\begin{align}
d(\bm{x}) = a \sin (2 \pi (x_1+1)/2) \sin (2 \pi (x_2+1)/2) (2 \pi (x_3+1)/2)
\end{align}
in each coordinate direction. An amplitude of~$a=0.15$ is used, leading to elements that are deformed significantly, see Figure~\ref{fig:geometries_and_meshes}. For the curvilinear mesh with deformation manifold, element mappings of polynomial degree~$3$ are used throughout this work independently of the polynomial degree of the shape functions.

\subsubsection{Robustness with respect to~$p$-refinement}\label{sec:PRobustnessCube}

\begin{table}[t]
\caption{Iteration count~$n_{10}$ as a function of polynomial degree~$p$ for various multigrid coarsening strategies for 3D \textbf{Cartesian} mesh with~$8^3$ elements. The smoother used for all experiments is Chebyshev(5,5) and the coarse-grid problem is solved iteratively to a relative tolerance of~$10^{-3}$ by the conjugate gradient method with AMG V-cycle as preconditioner.}
\label{tab:IterationCounts_3D_Cartesian_P_Robustness}
\renewcommand{\arraystretch}{1.1}
\begin{scriptsize}
\begin{center}
\subtable[$h$-multigrid (with~$p_{l-1} = \lfloor p_l/2 \rfloor $ if~$p$-transfer is involved)]{
\begin{tabular}{lrrrrrrrrrrrrrrr}
\cline{1-16}
 & \multicolumn{15}{c}{Polynomial degree~$p$}\\
\cline{2-16}
MG type & 1 & 2 & 3 & 4 & 5 & 6 & 7 & 8 & 9 & 10 & 11 & 12 & 13 & 14 & 15\\
\hline
$h$   & 14.8 & 12.5 & 12.4 & 12.2 & 14.4 & 14.8 & 17.2 & 17.5 & 19.5 & 19.1 & 22.3 & 22.4 & 24.3 & 24.5 & 26.2\\
$hp$  & 14.8 & 12.5 & 12.4 & 12.2 & 14.4 & 14.9 & 17.2 & 17.5 & 19.6 & 19.1 & 22.3 & 22.4 & 24.2 & 24.6 & 26.1\\
$hpc$ & 14.8 & 12.5 & 12.4 & 12.2 & 14.4 & 14.9 & 17.2 & 17.5 & 19.6 & 19.1 & 22.3 & 22.4 & 24.2 & 24.6 & 26.1\\
$ch$  & 7.5 & 5.5 & 5.2 & 5.1 & 5.2 & 5.1 & 5.5 & 5.6 & 6.6 & 6.6 & 7.8 & 7.8 & 8.7 & 8.8 & 9.8\\
$chp$ & 7.5 & 5.5 & 5.2 & 5.1 & 5.2 & 5.1 & 5.5 & 5.6 & 6.6 & 6.7 & 7.8 & 7.8 & 8.8 & 8.9 & 9.8\\
\hline
\end{tabular}}

\subtable[$p$-multigrid ($p_{l-1} = p_l-1$)]{
\begin{tabular}{lrrrrrrrrrrrrrrr}
\cline{1-16}
 & \multicolumn{15}{c}{Polynomial degree~$p$}\\
\cline{2-16}
MG type & 1 & 2 & 3 & 4 & 5 & 6 & 7 & 8 & 9 & 10 & 11 & 12 & 13 & 14 & 15\\
\hline
$p$   & 3.3 & 12.4 & 11.2 & 11.2 & 11.3 & 10.5 & 11.0 & 11.4 & 11.8 & 11.5 & 12.5 & 12.6 & 12.8 & 13.3 & 13.5\\
$ph$  & 14.8 & 12.5 & 11.3 & 11.3 & 11.3 & 10.7 & 10.9 & 11.5 & 11.8 & 11.6 & 12.5 & 12.8 & 13.3 & 13.6 & 13.8\\
$phc$ & 14.8 & 12.5 & 11.3 & 11.3 & 11.3 & 10.7 & 10.9 & 11.5 & 11.8 & 11.6 & 12.5 & 12.8 & 13.3 & 13.6 & 13.8\\
$cp$  & 7.4 & 5.5 & 5.1 & 4.9 & 4.8 & 5.0 & 4.7 & 4.7 & 4.6 & 4.7 & 4.6 & 4.7 & 4.6 & 4.8 & 4.9\\
$cph$ & 7.5 & 5.5 & 5.1 & 4.9 & 4.8 & 5.0 & 4.7 & 4.7 & 4.6 & 4.7 & 4.6 & 4.7 & 4.6 & 4.8 & 4.9\\
\hline
\end{tabular}}

\subtable[$p$-multigrid ($p_{l-1} = \lfloor p_l/2 \rfloor $)]{
\begin{tabular}{lrrrrrrrrrrrrrrr}
\cline{1-16}
 & \multicolumn{15}{c}{Polynomial degree~$p$}\\
\cline{2-16}
MG type & 1 & 2 & 3 & 4 & 5 & 6 & 7 & 8 & 9 & 10 & 11 & 12 & 13 & 14 & 15\\
\hline
$p$   & 3.3 & 12.4 & 13.0 & 11.9 & 14.2 & 14.1 & 15.9 & 15.4 & 17.9 & 16.9 & 20.1 & 19.4 & 21.3 & 22.3 & 24.3\\
$ph$  & 14.8 & 12.5 & 13.9 & 12.0 & 14.3 & 14.2 & 16.0 & 15.3 & 17.8 & 16.9 & 20.2 & 19.8 & 21.6 & 21.9 & 23.9\\
$phc$ & 14.8 & 12.5 & 13.9 & 12.0 & 14.3 & 14.2 & 16.0 & 15.3 & 17.8 & 16.9 & 20.2 & 19.8 & 21.6 & 21.9 & 23.9\\
$cp$  & 7.4 & 5.5 & 5.1 & 4.9 & 5.1 & 4.8 & 5.0 & 5.1 & 5.6 & 5.4 & 6.3 & 6.3 & 7.1 & 7.0 & 7.8\\
$cph$ & 7.5 & 5.5 & 5.1 & 4.9 & 5.1 & 4.8 & 5.0 & 5.1 & 5.6 & 5.4 & 6.3 & 6.3 & 7.1 & 7.0 & 7.8\\
\hline
\end{tabular}}

\subtable[$p$-multigrid ($p_{l-1} = 1 \, \forall p_l$)]{
\begin{tabular}{lrrrrrrrrrrrrrrr}
\cline{1-16}
 & \multicolumn{15}{c}{Polynomial degree~$p$}\\
\cline{2-16}
MG type & 1 & 2 & 3 & 4 & 5 & 6 & 7 & 8 & 9 & 10 & 11 & 12 & 13 & 14 & 15\\
\hline
$p$   & 3.3 & 12.4 & 13.0 & 16.6 & 20.3 & 24.6 & 29.4 & 35.6 & 41.6 & 44.9 & 55.8 & 64.6 & 74.0 & 86.9 & 96.8\\
$ph$  & 14.8 & 12.5 & 13.9 & 16.8 & 20.7 & 24.7 & 29.8 & 35.6 & 41.6 & 44.8 & 54.8 & 65.2 & 75.1 & 88.4 & 97.3\\
$phc$ & 14.8 & 12.5 & 13.9 & 16.8 & 20.7 & 24.7 & 29.8 & 35.6 & 41.6 & 44.8 & 54.7 & 65.2 & 75.2 & 88.6 & 97.0\\
$cp$  & 7.4 & 5.5 & 5.1 & 5.2 & 6.6 & 8.7 & 10.7 & 12.9 & 15.5 & 17.4 & 19.9 & 22.6 & 24.7 & 26.9 & 29.8\\
$cph$ & 7.5 & 5.5 & 5.1 & 5.2 & 6.6 & 8.7 & 10.7 & 12.9 & 15.5 & 17.4 & 19.9 & 22.5 & 24.6 & 26.9 & 29.7\\
\hline
\end{tabular}}

\end{center}
\end{scriptsize}
\renewcommand{\arraystretch}{1}
\end{table}

\begin{table}[t]
\caption{Iteration count~$n_{10}$ as a function of polynomial degree~$p$ for various multigrid coarsening strategies for 3D \textbf{curvilinear} mesh with~$8^3$ elements. The smoother used for all experiments is Chebyshev(5,5) and the coarse-grid problem is solved iteratively to a relative tolerance of~$10^{-3}$ by the conjugate gradient method with AMG V-cycle as preconditioner.}
\label{tab:IterationCounts_3D_Curvilinear_P_Robustness}
\renewcommand{\arraystretch}{1.1}
\begin{scriptsize}
\begin{center}
\subtable[$h$-multigrid (with~$p_{l-1} = \lfloor p_l/2 \rfloor $ if~$p$-transfer is involved)]{
\begin{tabular}{lrrrrrrrrrrrrrrr}
\cline{1-16}
 & \multicolumn{15}{c}{Polynomial degree~$p$}\\
\cline{2-16}
MG type & 1 & 2 & 3 & 4 & 5 & 6 & 7 & 8 & 9 & 10 & 11 & 12 & 13 & 14 & 15\\
\hline
$h$   & 17.7 & 13.8 & 14.0 & 15.0 & 17.8 & 19.4 & 22.6 & 22.6 & 25.4 & 26.4 & 28.9 & 29.8 & 32.3 & 33.3 & 35.8\\
$hp$  & 17.7 & 13.8 & 14.0 & 15.0 & 17.8 & 19.4 & 22.6 & 23.1 & 25.3 & 26.3 & 28.8 & 29.7 & 32.3 & 32.9 & 35.5\\
$hpc$ & 17.7 & 13.8 & 14.0 & 15.0 & 17.8 & 19.4 & 22.6 & 23.1 & 25.3 & 26.3 & 28.8 & 29.7 & 32.3 & 32.9 & 35.5\\
$ch$  & 8.5 & 5.9 & 5.5 & 5.5 & 5.9 & 6.5 & 8.1 & 8.7 & 10.2 & 10.8 & 12.2 & 12.7 & 13.9 & 14.4 & 15.7\\
$chp$ & 8.5 & 5.9 & 5.5 & 5.5 & 5.9 & 6.6 & 7.9 & 8.7 & 10.1 & 10.7 & 12.1 & 12.6 & 13.9 & 14.4 & 15.7\\
\hline
\end{tabular}}

\subtable[$p$-multigrid ($p_{l-1} = p_l-1$)]{
\begin{tabular}{lrrrrrrrrrrrrrrr}
\cline{1-16}
 & \multicolumn{15}{c}{Polynomial degree~$p$}\\
\cline{2-16}
MG type & 1 & 2 & 3 & 4 & 5 & 6 & 7 & 8 & 9 & 10 & 11 & 12 & 13 & 14 & 15\\
\hline
$p$   & 3.4 & 13.7 & 12.8 & 13.0 & 13.8 & 15.0 & 15.5 & 16.3 & 16.6 & 16.7 & 17.7 & 17.9 & 18.7 & 18.8 & 19.6\\
$ph$  & 17.7 & 14.2 & 13.2 & 13.1 & 13.9 & 15.1 & 15.5 & 16.3 & 16.7 & 16.8 & 17.8 & 18.0 & 18.7 & 18.7 & 19.7\\
$phc$ & 17.7 & 14.2 & 13.2 & 13.1 & 13.9 & 15.1 & 15.5 & 16.3 & 16.7 & 16.8 & 17.8 & 18.0 & 18.7 & 18.7 & 19.7\\
$cp$  & 8.5 & 5.9 & 5.4 & 5.3 & 5.3 & 5.2 & 5.2 & 5.3 & 5.8 & 5.9 & 6.5 & 6.7 & 7.4 & 7.6 & 8.0\\
$cph$ & 8.5 & 5.9 & 5.4 & 5.3 & 5.3 & 5.2 & 5.2 & 5.3 & 5.8 & 5.9 & 6.5 & 6.7 & 7.4 & 7.6 & 7.9\\
\hline
\end{tabular}}

\subtable[$p$-multigrid ($p_{l-1} = \lfloor p_l/2 \rfloor $)]{
\begin{tabular}{lrrrrrrrrrrrrrrr}
\cline{1-16}
 & \multicolumn{15}{c}{Polynomial degree~$p$}\\
\cline{2-16}
MG type & 1 & 2 & 3 & 4 & 5 & 6 & 7 & 8 & 9 & 10 & 11 & 12 & 13 & 14 & 15\\
\hline
$p$   & 3.4 & 13.7 & 15.3 & 14.0 & 18.5 & 19.5 & 22.2 & 21.7 & 25.1 & 25.9 & 29.1 & 29.2 & 33.0 & 33.8 & 36.5\\
$ph$  & 17.7 & 14.2 & 16.4 & 14.1 & 18.6 & 19.5 & 22.3 & 21.9 & 25.1 & 25.9 & 29.1 & 29.3 & 32.8 & 33.7 & 36.6\\
$phc$ & 17.7 & 14.2 & 16.4 & 14.1 & 18.6 & 19.5 & 22.3 & 21.9 & 25.1 & 25.9 & 29.1 & 29.3 & 32.8 & 33.7 & 36.6\\
$cp$  & 8.5 & 5.9 & 5.8 & 5.5 & 6.5 & 6.3 & 7.8 & 7.8 & 9.5 & 9.7 & 10.9 & 11.3 & 12.6 & 12.7 & 13.9\\
$cph$ & 8.5 & 5.9 & 5.9 & 5.5 & 6.5 & 6.3 & 7.8 & 7.8 & 9.5 & 9.7 & 10.9 & 11.3 & 12.6 & 12.7 & 13.9\\
\hline
\end{tabular}}

\subtable[$p$-multigrid ($p_{l-1} = 1 \, \forall p_l$)]{
\begin{tabular}{lrrrrrrrrrrrrrrr}
\cline{1-16}
 & \multicolumn{15}{c}{Polynomial degree~$p$}\\
\cline{2-16}
MG type & 1 & 2 & 3 & 4 & 5 & 6 & 7 & 8 & 9 & 10 & 11 & 12 & 13 & 14 & 15\\
\hline
$p$   & 3.4 & 13.7 & 15.3 & 20.3 & 27.0 & 34.7 & 39.2 & 45.9 & 54.0 & 63.7 & 76.4 & 94.9 & 115 & 137 & 157\\
$ph$  & 17.7 & 14.2 & 16.4 & 21.0 & 27.0 & 34.4 & 39.1 & 45.9 & 53.4 & 62.9 & 75.8 & 94.0 & 113 & 135 & 156\\
$phc$ & 17.7 & 14.2 & 16.4 & 21.0 & 27.0 & 34.4 & 39.1 & 45.9 & 53.4 & 62.9 & 75.8 & 93.6 & 113 & 135 & 156\\
$cp$  & 8.5 & 5.9 & 5.8 & 8.4 & 11.7 & 15.2 & 18.9 & 23.0 & 27.1 & 31.4 & 36.0 & 40.4 & 45.0 & 49.6 & 54.2\\
$cph$ & 8.5 & 5.9 & 5.9 & 8.4 & 11.7 & 15.2 & 18.9 & 23.0 & 27.0 & 31.4 & 35.9 & 40.3 & 44.9 & 49.5 & 54.2\\
\hline
\end{tabular}}

\end{center}
\end{scriptsize}
\renewcommand{\arraystretch}{1}
\end{table}

In a first numerical experiment, we investigate the number of iterations as a function of the polynomial degree~$p$ for various multigrid coarsening strategies discussed in Section~\ref{sec:HybridMultigrid}. Table~\ref{tab:SummaryCoarseningStrategies} summarizes all possible multigrid coarsening types. We distinguish between~$h$-like MG approaches where additional~$p$-coarsening is done on the coarsest~$h$-level~($hp$-MG), and~$p$-like MG approaches where additional~$h$-coarsening is done at lowest degree~$p=1$~($ph$-MG). Table~\ref{tab:IterationCounts_3D_Cartesian_P_Robustness} lists the results obtained for the Cartesian mesh and Table~\ref{tab:IterationCounts_3D_Curvilinear_P_Robustness} the results obtained for the curvilinear mesh. While we consider the three different~$p$-coarsening strategies from Section~\ref{sec:HybridMultigrid} for the~$p$-like approaches, the~$h$-like approaches exclusively use the~$p$-coarsening denoted as bisection that approximately halves the number of unknowns per direction from one multigrid level to the next. With respect to additional~$c$-coarsening, we do not explicitly list all possible combinations in Tables~\ref{tab:IterationCounts_3D_Cartesian_P_Robustness} and~\ref{tab:IterationCounts_3D_Curvilinear_P_Robustness}, but focus on those that we consider most important or interesting and comment on the remaining ones in the text. A fixed number of elements of~$8^3$ is used and the polynomial degree varies between~$1\leq p \leq 15$. The results can be summarized as follows:

\begin{itemize}
\item Extending the pure~$h$- or~$p$-multigrid methods towards hybrid multigrid methods with additional~$p$- or~$h$-coarsening, respectively, on coarser levels does not change the multigrid convergence rates. The multigrid convergence rates are also not altered if additional~$c$-coarsening is performed at the coarsest level before the coarse-grid solver is invoked ($hc$-,~$hpc$- and~$pc$,~$phc$-approaches) or at an intermediate level between~$h$- and~$p$-coarsening ($hcp$- and~$pch$-approaches).
\item A different convergence behavior with much lower iteration counts is observed when performing the~$c$-transfer on the finest level before~$h$- or~$p$-coarsening is invoked. For all multigrid approaches and for both Cartesian and curvilinear meshes, iteration counts are reduced by a factor of~$2$ to~$3$ compared to~$c$-coarsening performed on coarser levels. Performing the~$c$-transfer introduces additional costs as quantified in Section~\ref{sec:OptimalMultigridSequence}.
\item With respect to~$p$-robustness, the~$h$-like approaches on the one hand and the~$p$-like approaches with~$p_{l-1} = \lfloor p_l/2 \rfloor $ coarsening on the other hand show a similar relation between polynomial degree and iteration counts. This is not unexpected, as both approaches reduce the degrees of freedom in factors of two per direction per level. In combination with the Chebyshev smoother considered here, these approaches show a slight increase in iteration counts for large~$p$.
\item The~$p$-multigrid methods with~$p_{l-1} = 1 \, \forall p_l$ coarsening show a much stronger increase in iteration counts for large~$p$. The results shown here also shed light on previous results~\cite{Siefert2014, Bastian2019, Remacle2016}, where two-level approaches with an immediate transfer from highest to lowest polynomial degree have been used. We will show in the following that this coarsening strategy is not only performing worst in terms of iteration counts, but also in terms of computational costs.
\item The~$p$-multigrid methods with~$p_{l-1} = p_l-1$ coarsening show the best behavior in terms of~$p$-robustness w.r.t.~iteration counts. On the Cartesian mesh, the iteration counts are completely independent of~$p$ for the~$cp$- and~$cph$-approaches, and the number of iterations increases only slightly for increasing~$p$ on the curvilinear mesh. However, this type of~$p$-coarsening is also the most complex one introducing the largest numbers of multigrid levels. Hence, from the results shown in Tables~\ref{tab:IterationCounts_3D_Cartesian_P_Robustness} and~\ref{tab:IterationCounts_3D_Curvilinear_P_Robustness}, it is unclear whether this strategy pays off in terms of computational costs, an aspect investigated in detail in Section~\ref{sec:OptimalMultigridSequence}.
\end{itemize}

\subsubsection{Robustness with respect to~$h$-refinement}

\begin{table}[!ht]
\caption{Robustness of multigrid algorithm with respect to mesh size~$h$ for polynomial degrees~$p=1,\ldots, 15$ and different multigrid coarsening strategies. The table lists the iteration count~$n_{10}$. The considered test case is the cube test case on a 3D Cartesian mesh with~$4^3, 8^3, 16^3$ elements. The smoother used for all experiments is Chebyshev(5,5) and the coarse-grid problem is solved iteratively to a relative tolerance of~$10^{-3}$ by the conjugate gradient method with AMG V-cycle as preconditioner.}
\label{tab:IterationCounts_3D_Cartesian_H_Robustness}
\renewcommand{\arraystretch}{1.1}
\begin{scriptsize}
\begin{center}
\subtable[pure $h$-multigrid]{
\begin{tabular}{lrrrrrrrrrrrrrrr}
\cline{1-16}
 & \multicolumn{15}{c}{Polynomial degree~$p$}\\
\cline{2-16}
$h/L$ & 1 & 2 & 3 & 4 & 5 & 6 & 7 & 8 & 9 & 10 & 11 & 12 & 13 & 14 & 15\\
\hline
$1/4$  & 8.0  & 9.8  & 11.2 & 12.0 & 13.3 & 14.1 & 15.2 & 16.5 & 18.1 & 18.3 & 20.3 & 21.1 & 22.4 & 22.9 & 24.4\\
$1/8$  & 14.8 & 12.5 & 12.4 & 12.2 & 14.4 & 14.8 & 17.2 & 17.5 & 19.5 & 19.1 & 22.3 & 22.4 & 24.3 & 24.5 & 26.2\\
$1/16$ & 16.8 & 13.3 & 12.6 & 12.7 & 14.9 & 14.9 & 17.1 & 17.7 & 19.9 & 19.2 & 22.6 & 22.8 & 24.8 & 24.9 & 26.7\\
\hline
\end{tabular}}

\subtable[pure $p$-multigrid ($p_{l-1} = \lfloor p_l/2 \rfloor $)]{
\begin{tabular}{lrrrrrrrrrrrrrrr}
\cline{1-16}
 & \multicolumn{15}{c}{Polynomial degree~$p$}\\
\cline{2-16}
$h/L$ & 1 & 2 & 3 & 4 & 5 & 6 & 7 & 8 & 9 & 10 & 11 & 12 & 13 & 14 & 15\\
\hline
$1/4$  & 1.5 & 9.3  & 12.2 & 11.5 & 13.5 & 13.6 & 15.8 & 16.0 & 18.4 & 17.8 & 20.5 & 19.9 & 22.2 & 21.6 & 23.9\\
$1/8$  & 3.3 & 12.4 & 13.0 & 11.9 & 14.2 & 14.1 & 15.9 & 15.4 & 17.9 & 16.9 & 20.1 & 19.4 & 21.3 & 22.3 & 24.3\\
$1/16$ & 3.4 & 13.0 & 12.8 & 12.1 & 14.0 & 14.2 & 16.0 & 15.2 & 17.7 & 15.7 & 19.0 & 19.6 & 21.9 & 22.5 & 23.8\\
\hline
\end{tabular}}

\subtable[$cph$-multigrid ($p_{l-1} = \lfloor p_l/2 \rfloor $)]{
\begin{tabular}{lrrrrrrrrrrrrrrr}
\cline{1-16}
 & \multicolumn{15}{c}{Polynomial degree~$p$}\\
\cline{2-16}
$h/L$ & 1 & 2 & 3 & 4 & 5 & 6 & 7 & 8 & 9 & 10 & 11 & 12 & 13 & 14 & 15\\
\hline
$1/4$  & 5.7 & 5.6 & 5.3 & 4.9 & 5.1 & 4.8 & 5.2 & 4.8 & 5.2 & 5.2 & 5.9 & 6.2 & 6.9 & 6.9 & 7.7\\
$1/8$  & 7.5 & 5.5 & 5.1 & 4.9 & 5.1 & 4.8 & 5.0 & 5.1 & 5.6 & 5.4 & 6.3 & 6.3 & 7.1 & 7.0 & 7.8\\
$1/16$ & 7.4 & 5.4 & 5.5 & 5.1 & 5.2 & 5.1 & 5.3 & 5.0 & 5.6 & 5.5 & 6.4 & 6.4 & 7.2 & 7.3 & 7.8\\
\hline
\end{tabular}}


\end{center}
\end{scriptsize}
\renewcommand{\arraystretch}{1}
\end{table}

Results of~$h$-robustness tests are shown in Table~\ref{tab:IterationCounts_3D_Cartesian_H_Robustness}. As representative multigrid methods, we selected the pure~$h$- and~$p$-multigrid methods and the combined~$cph$-multigrid method, each of them showing mesh independent convergence as expected. Robustness with respect to~$h$-refinement is also achieved for the~$chp$-coarsening strategy (not shown explicitly in Table~\ref{tab:IterationCounts_3D_Cartesian_H_Robustness}) with iteration counts slightly larger than for the~$cph$-multigrid method, in agreement with the results in Table~\ref{tab:IterationCounts_3D_Cartesian_P_Robustness}. The~$cph$-coarsening strategy is shown here as a representative hybrid multigrid method for reasons of computational efficiency, as explained below in Section~\ref{sec:OptimalMultigridSequence}, where this method is identified as a very efficient coarsening strategy. Similarly, we also obtained~$h$-robustness for the 3D curvilinear mesh, but omit these results here for reasons of brevity.

\begin{table}[!h]
\caption{Robustness of multigrid algorithm with respect to interior penalty parameter. The table lists the iteration count~$n_{10}$. The considered test case is the cube test case on a 3D Cartesian mesh with~$8^3$ elements. The smoother used for all experiments is Chebyshev(5,5) and the coarse-grid problem is solved iteratively to a relative tolerance of~$10^{-3}$ by the conjugate gradient method with AMG V-cycle as preconditioner.}
\label{tab:IterationCounts_3D_Cartesian_PenaltyFactor}
\renewcommand{\arraystretch}{1.1}
\begin{scriptsize}

\begin{center}
\subtable[$hp$-multigrid ($p_{l-1} = \lfloor p_l/2 \rfloor $)]{
\begin{tabular}{lrrrrrrrrrrrrrrr}
\cline{1-16}
 & \multicolumn{15}{c}{Polynomial degree~$p$}\\
\cline{2-16}
IP factor & 1 & 2 & 3 & 4 & 5 & 6 & 7 & 8 & 9 & 10 & 11 & 12 & 13 & 14 & 15\\
\hline
$10^0 \cdot \tau$ & 14.8 & 12.5 & 12.4 & 12.2 & 14.4 & 14.9 & 17.2 & 17.5 & 19.6 & 19.1 & 22.3 & 22.4 & 24.2 & 24.6 & 26.1\\
$10^1 \cdot \tau$ & 25.4 & 32.6 & 39.9 & 39.7 & 46.8 & 45.4 & 52.6 & 51.8 & 55.8 & 56.5 & 62.0 & 62.3 & 67.9 & 68.5 & 73.0\\
$10^2 \cdot \tau$ & 38.5 & 53.8 & 79.9 & 83.7 & 109  & 104  & 128  & 117  & 134  & 133  & 147  & 146  & 157  & 166  & 176 \\
$10^3 \cdot \tau$ & 45.0 & 73.3 & 113  & 123  & 172  & 162  & 205  & 190  & 223  & 194  & 221  & 219  & 243  & 249  & 278 \\
\hline
\end{tabular}}

\subtable[$ph$-multigrid ($p_{l-1} = \lfloor p_l/2 \rfloor $)]{
\begin{tabular}{lrrrrrrrrrrrrrrr}
\cline{1-16}
 & \multicolumn{15}{c}{Polynomial degree~$p$}\\
\cline{2-16}
IP factor & 1 & 2 & 3 & 4 & 5 & 6 & 7 & 8 & 9 & 10 & 11 & 12 & 13 & 14 & 15\\
\hline
$10^0 \cdot \tau$ & 14.8 & 12.5 & 13.9 & 12.0 & 14.3 & 14.2 & 16.0 & 15.3 & 17.8 & 16.9 & 20.2 & 19.8 & 21.6 & 21.9 & 23.9\\
$10^1 \cdot \tau$ & 25.4 & 29.8 & 33.8 & 33.2 & 38.2 & 38.4 & 43.2 & 42.6 & 48.7 & 46.4 & 52.3 & 52.2 & 56.7 & 57.7 & 61.3\\
$10^2 \cdot \tau$ & 38.5 & 45.2 & 49.6 & 52.2 & 61.7 & 63.4 & 67.5 & 70.7 & 80.0 & 77.8 & 85.7 & 85.5 & 96.4 & 93.3 & 102 \\
$10^3 \cdot \tau$ & 45.0 & 59.1 & 66.0 & 70.9 & 79.4 & 80.1 & 89.0 & 94.8 & 108  &  105 & 116  & 119  & 126  & 125  & 138 \\
\hline
\end{tabular}}

\subtable[$cph$-multigrid ($p_{l-1} = \lfloor p_l/2 \rfloor $)]{
\begin{tabular}{lrrrrrrrrrrrrrrr}
\cline{1-16}
 & \multicolumn{15}{c}{Polynomial degree~$p$}\\
\cline{2-16}
IP factor & 1 & 2 & 3 & 4 & 5 & 6 & 7 & 8 & 9 & 10 & 11 & 12 & 13 & 14 & 15\\
\hline
$10^0 \cdot \tau$ & 7.5 & 5.5 & 5.1 & 4.9 & 5.1 & 4.8 & 5.0 & 5.1 & 5.6 & 5.4 & 6.3 & 6.3 & 7.1 & 7.0 & 7.8\\
$10^1 \cdot \tau$ & 7.7 & 5.4 & 5.3 & 5.3 & 5.4 & 5.2 & 5.3 & 5.6 & 5.7 & 5.7 & 6.4 & 6.5 & 7.4 & 7.2 & 8.0\\
$10^2 \cdot \tau$ & 7.7 & 5.4 & 5.3 & 5.4 & 5.5 & 5.4 & 5.4 & 5.7 & 5.8 & 5.8 & 6.5 & 6.5 & 7.2 & 7.2 & 8.1\\
$10^3 \cdot \tau$ & 7.7 & 5.4 & 5.4 & 5.4 & 5.5 & 5.4 & 5.4 & 5.7 & 5.9 & 5.9 & 6.9 & 6.8 & 7.6 & 7.8 & 8.8\\
\hline
\end{tabular}}


\end{center}
\end{scriptsize}
\renewcommand{\arraystretch}{1}
\end{table}

\subsubsection{Robustness with respect to interior penalty parameter}

The coarsening strategies performing the~$c$-transfer on the finest level (such as~$ch$-,~$cp$-,~$cph$-,~$chp$-coarsening) have the interesting property that the resulting multigrid algorithm exhibits convergence rates that are independent of the penalty factor of the interior penalty method. This property is demonstrated in Table~\ref{tab:IterationCounts_3D_Cartesian_PenaltyFactor}, where the~$cph$-multigrid method is compared to combined~$hp$- and~$ph$-multigrid methods without~$c$-transfer (pure~$p$- and pure~$h$-multigrid methods would show a qualitatively similar behavior). As expected, for standard~$hp$- and~$ph$-coarsening, the iteration counts degrade significantly when increasing the interior penalty factor, while the~$cph$-multigrid method shows constant iteration counts when scaling the penalty factor by~$10^0, 10^1, 10^2, 10^3$. The~$chp$-multigrid approach also shows robustness with respect to the interior penalty parameter~$\tau$, and is not shown explicitly in Table~\ref{tab:IterationCounts_3D_Cartesian_PenaltyFactor} for the sake of brevity. Qualitatively, we obtained the same results when repeating this experiment for the 3D curvilinear mesh. An explanation for this~$\tau$-robustness might be that the continuous finite element space covers the DG solution in the limit of large penalty factors,  thereby balancing the deteriorating conditioning of the DG operator, see also the theory in~\cite{Antonietti2017}. In other words, the interior penalty parameter does not only impact the conditioning, but also the approximation properties of the DG solution in relation to the continuous FE space. This behavior is appealing as it allows to avoid the need to optimize the IP parameter in order to obtain iteration counts as low as possible and, at the same time, ensure coercivity of the IP method.

\subsubsection{Identification of optimal multigrid sequence maximizing throughput}\label{sec:OptimalMultigridSequence}
The results in Section~\ref{sec:PRobustnessCube} revealed that using a larger number of~$p$-levels in the multigrid hierarchy reduces the number of iterations, at the costs of increased computational load per iteration. However, it remains unclear which type of~$p$-coarsening is the most efficient one. Likewise, it needs to be investigated whether a~$c$-transfer at the finest level (with an additional expensive smoothing step performed on the finest level as opposed to a cheap~$c$-transfer at an intermediate level or at the coarsest level) reduces overall computational costs. As mentioned in the introduction, the driving force for algorithmic selections should be time-to-solution, and we address these questions in this section using the performance metrics introduced in Section~\ref{sec:Metrics}.

\begin{figure}[!ht]
\centering
\includegraphics[width=0.85\textwidth]{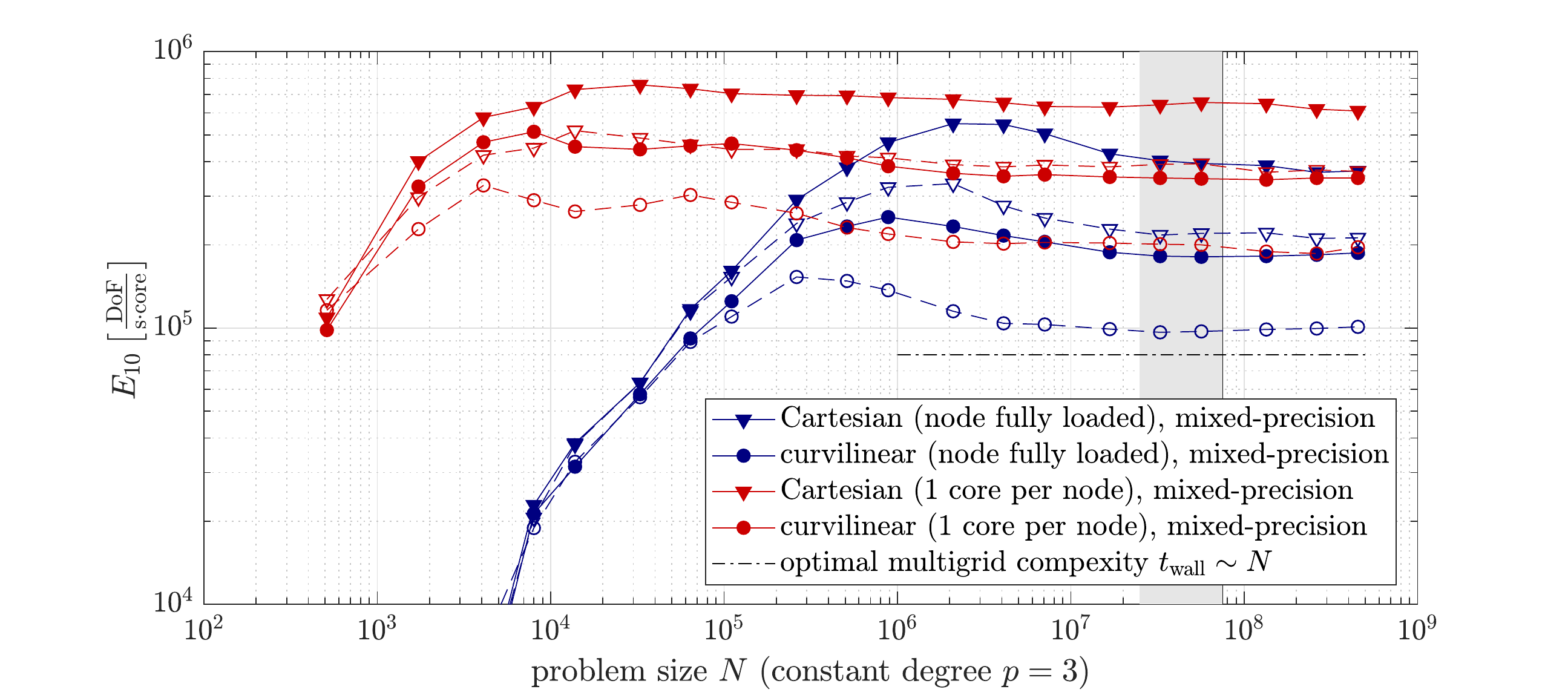}
\caption{Throughput~$E_{10}$ versus problem size for polynomial degree~$p=3$ and cube test case with Cartesian and curvilinear meshes. A~$cph$-multigrid coarsening strategy is used with~$p_{l-1} = \lfloor p_l/2 \rfloor$. Standard mixed-precision multigrid results are shown as solid lines, and additional computations performed in double precision only are shown as dashed lines. The gray band indicates the range of problem sizes used for the throughput measurements in Figure~\ref{fig:sine_iterations_vs_throughput}, for which a fully loaded node (blue curves) is considered with the problem size large enough to saturate caches. A fat memory node is used here in order to investigate a wide range of problem sizes.}
\label{fig:throuhgput_cube_1core_vs_1node}
\end{figure}

We note that for throughput measurements it is important to fully utilize all cores of one compute node since certain resources are shared by the cores of a node, i.e., the performance reported in degrees of freedom solved per second per core would otherwise be extraordinarily high. This is demonstrated in Figure~\ref{fig:throuhgput_cube_1core_vs_1node}, where the throughput is significantly larger if only a single core is utilized per node instead of a fully loaded node.  Figure~\ref{fig:throuhgput_cube_1core_vs_1node} also shows the speed-up that can be achieved by the use of mixed-precision multigrid, which is around a factor of~$1.8$ for large problem sizes. Towards very small problem sizes (strong scaling limit), the performance breaks down since performance is limited by latency and the available parallelism instead of arithmetic throughput or memory throughput, and the performance advantage of mixed-precision multigrid therefore vanishes in such a scenario. For the computations on a fully-loaded node, an elevation of the throughput can be observed for problem sizes around~$1\ \mathrm{MDoF}$ due to the fact that data fits (partly) into caches, which have a higher bandwidth than main memory. We therefore run throughput measurements in a saturated regime of sufficiently high workload per core so that the data does no longer fit into caches. In Figure~\ref{fig:throuhgput_cube_1core_vs_1node}, the throughput is shown as a function of the problem size to highlight these cache effects and we indicate the range of problem sizes ($25\mathrm{MDoF}-75 \mathrm{MDoF}$) used below for benchmarking the present solver by a gray band. While it is good practice to run  the solver in a saturated regime for benchmarking, it is of course beneficial to explicitly exploit caching effects for practical computations.

\begin{figure}[!ht]
\centering
 \subfigure[Cube test case on 3D Cartesian mesh.]{
\includegraphics[width=1.0\textwidth]{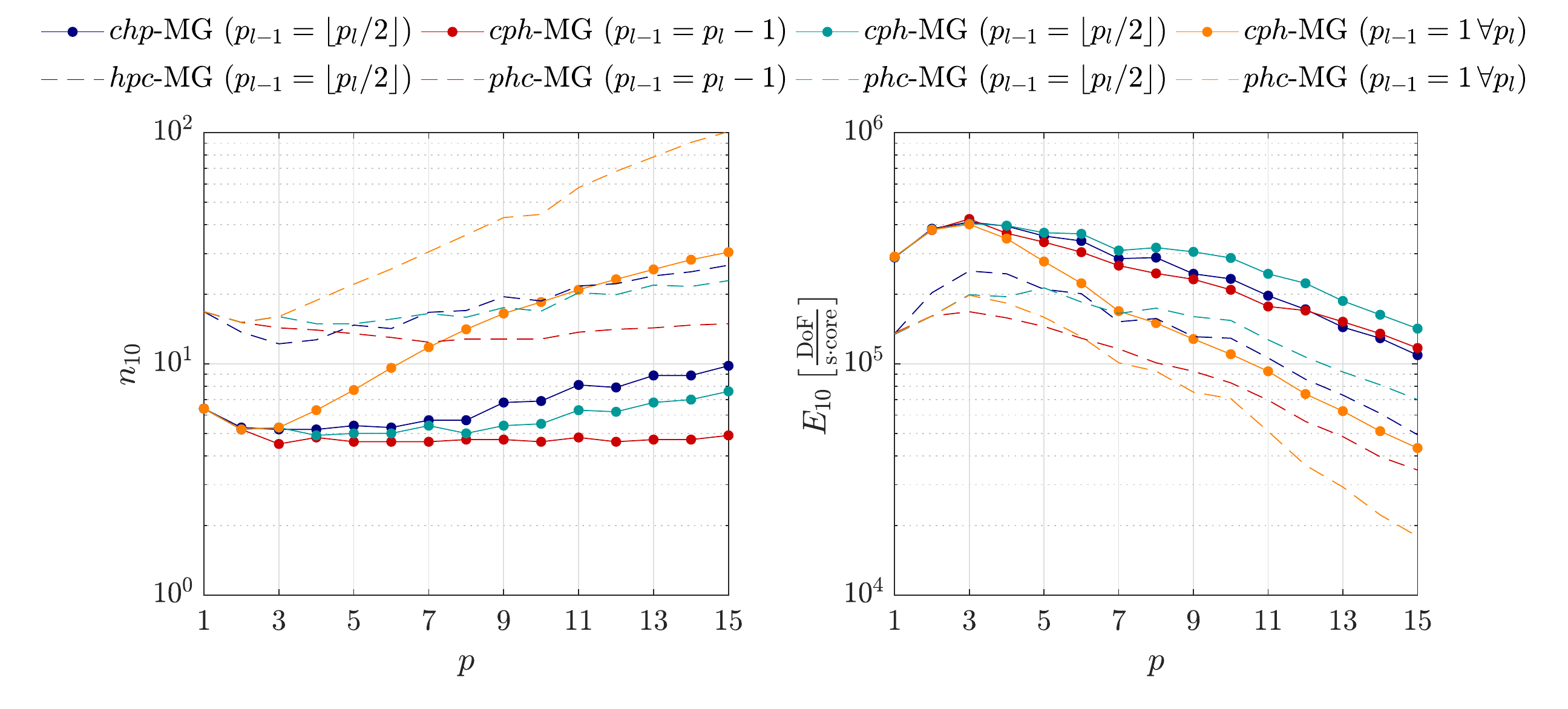}}
 \subfigure[Cube test case on 3D curvilinear mesh.]{
\includegraphics[width=1.0\textwidth]{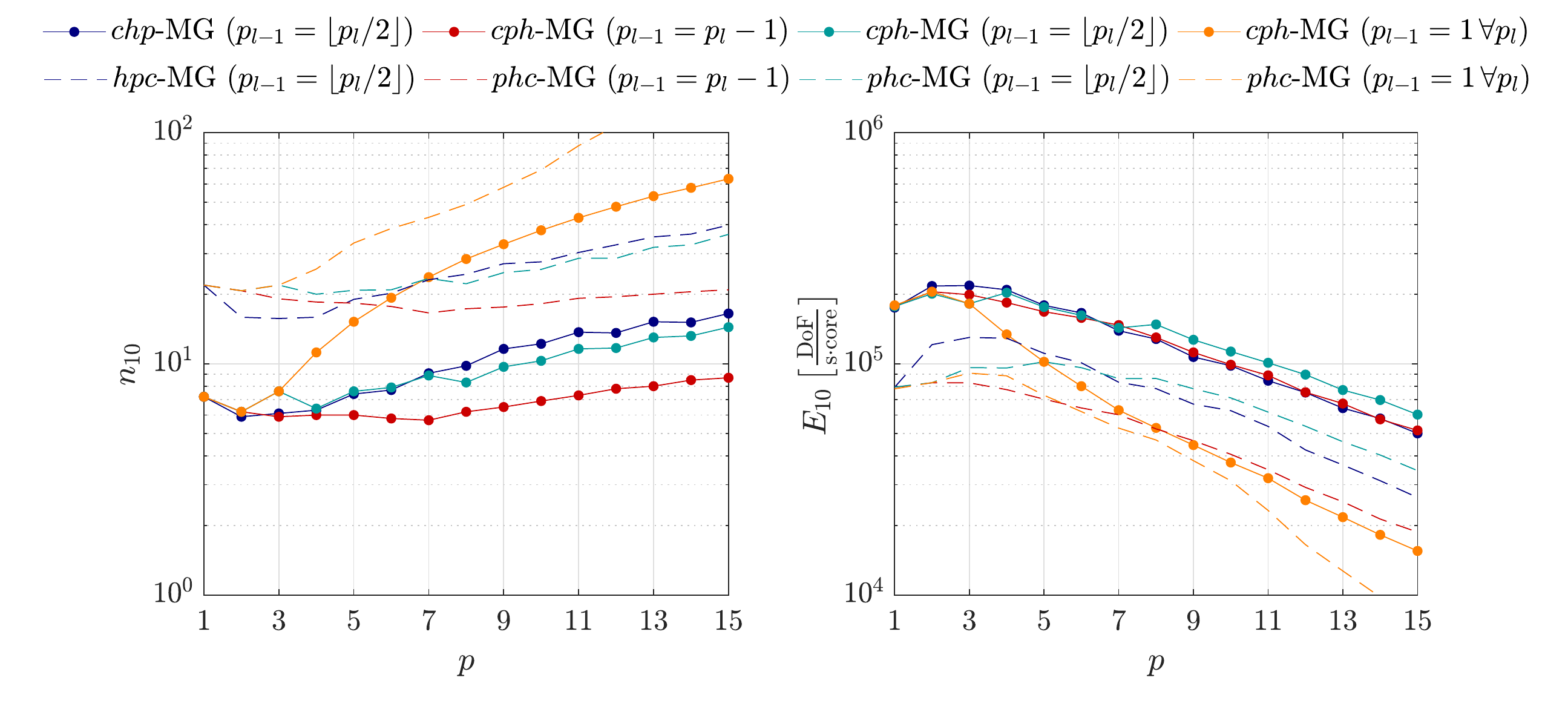}}
\caption{Iterations~$n_{10}$ versus throughput~$E_{10}$ for different multigrid coarsening strategies on Cartesian mesh and curvilinear mesh. The smoother used for all experiments is Chebyshev(5,5) and the coarse-grid problem is solved iteratively to a relative tolerance of~$10^{-3}$ by the conjugate gradient method with AMG V-cycle as preconditioner. The problem size is between~$25\mathrm{MDoF}-75 \mathrm{MDoF}$ for all polynomial degrees~$1\leq p \leq 15$.}
\label{fig:sine_iterations_vs_throughput}
\end{figure}

In Figure~\ref{fig:sine_iterations_vs_throughput}, we detail the performance in terms of iteration counts as well as computational efficiency for different hybrid multigrid algorithms, all of them exploiting all levels of~$h$-,~$p$-, and~$c$-coarsening (in different orders). For the~$p$-like approaches, we again investigate the three different types of~$p$-coarsening. The results for~$n_{10}$ in the left panels of the figures visualize results similar to those shown in Tables~\ref{tab:IterationCounts_3D_Cartesian_P_Robustness} and~\ref{tab:IterationCounts_3D_Curvilinear_P_Robustness}. The~$cph$- and~$phc$-methods with~$p_{l-1} = p_l-1$ coarsening exhibit a constant number of iterations for large~$p$ on the Cartesian mesh, and a slight increase in the number of iterations for the curvilinear mesh. The results in Figure~\ref{fig:sine_iterations_vs_throughput} highlight that performing the~$c$-transfer on the finest level is not only beneficial in terms of iteration counts, but also in terms of computational costs. The~$chp$- and~$cph$-multigrid methods outperform the~$hpc$- and~$phc$-multigrid methods on the Cartesian mesh as well as on the curvilinear mesh.~\footnote{Note that the difference would be smaller if the interior penalty parameter would be chosen as close to the minimal value ensuring coercivity as possible, reducing~$n_{10}$ for the~$hpc$- and~$phc$-multigrid methods. At the same time, the~$chp$- and~$cph$-approaches will be significantly faster than the~$hpc$- and~$phc$-multigrid methods for larger penalty factors, see the results in Table~\ref{tab:IterationCounts_3D_Cartesian_PenaltyFactor}.} In terms of~$p$-coarsening, the~$p_{l-1} = 1 \, \forall p_l$ strategy performs worst both in terms of iteration counts and computational costs. The~$p_{l-1} = \lfloor p_l/2 \rfloor $ strategy performs best in terms of computational costs. The differences between~$hp$- versus~$ph$-multigrid methods are very small, and small differences in the number of iterations determine which approach is more efficient overall. Despite exhibiting the lowest number of iterations, the~$p_{l-1} = p_l-1$ strategy is not competitive if the~$c$-transfer is done at the coarse level. However, it is interesting to realize that the~$p_{l-1} = p_l-1$ strategy can keep up with the~$p_{l-1} = \lfloor p_l/2 \rfloor $ coarsening strategy if the~$c$-transfer is performed on the fine level. Going through all polynomial degrees in the multigrid hierarchy introduces less overhead in this case since the operator evaluation is significantly faster for the continuous FE space (e.g., no face integrals) compared to the DG space~\cite{Kronbichler2018}. Although not explicitly shown here, it should be mentioned that the increased number of multigrid levels for the~$p_{l-1} = p_l-1$ strategy is disadvantageous due to increased memory requirements, and also in the strong scaling limit where overall costs are dominated by the latency of matrix-vector products (and hence the number of multigrid levels). For this reason, we consider the~$chp$- and~$cph$-multigrid methods with~$p_{l-1} = \lfloor p_l/2 \rfloor $ coarsening strategy the most promising methods that are investigated below for the more challenging test cases. 

\begin{table}[!h]
\caption{Iteration count~$n_{10}$ and effective number of fine-level matrix-vector products~$n_{10, \mathrm{mat-vec}}$ for Cartesian mesh versus curvilinear mesh in 3D. The~$cph$-multigrid method with~$p_{l-1} = \lfloor p_l/2 \rfloor$ and Chebyshev(5,5) smoother is used. The problem size is between~$25\mathrm{MDoF}-75 \mathrm{MDoF}$ for all polynomial degrees~$1\leq p \leq 15$.}
\label{tab:Throughput_3D_Cartesian_vs_Curvilinear}
\renewcommand{\arraystretch}{1.1}
\begin{scriptsize}
\begin{center}

\begin{tabular}{cccccccccc}
\cline{1-10}
 & \multicolumn{4}{c}{3D Cartesian mesh} & & \multicolumn{4}{c}{3D curvilinear mesh}\\
\cline{2-5} \cline{7-10}
$p$ & $n_{10}$ & $n_{10, \mathrm{mat-vec}}$ & $E_{\mathrm{mat-vec}} \left[\frac{\mathrm{MDoF}}{\mathrm{s} \cdot \mathrm{core}}\right]$ & $E_{10} \left[\frac{\mathrm{MDoF}}{\mathrm{s} \cdot \mathrm{core}}\right]$ & & $n_{10}$ & $n_{10, \mathrm{mat-vec}}$ & $E_{\mathrm{mat-vec}} \left[\frac{\mathrm{MDoF}}{\mathrm{s} \cdot \mathrm{core}}\right]$ & $E_{10} \left[\frac{\mathrm{MDoF}}{\mathrm{s} \cdot \mathrm{core}}\right]$\\
\hline
1  & 6.4 & 90  & 26.0 & 0.289 & & 7.2  & 94  & 16.7 & 0.177\\
2  & 5.2 & 102 & 39.0 & 0.381 & & 6.2  & 99  & 19.8 & 0.201\\
3  & 5.3 & 97  & 39.0 & 0.404 & & 7.6  & 109 & 19.9 & 0.182\\
4  & 4.9 & 95  & 37.7 & 0.396 & & 6.4  & 98  & 20.0 & 0.203\\
5  & 5.0 & 106 & 39.0 & 0.369 & & 7.6  & 119 & 21.0 & 0.176\\
6  & 5.0 & 94  & 34.3 & 0.365 & & 7.9  & 122 & 19.7 & 0.162\\
7  & 5.4 & 99  & 30.5 & 0.309 & & 8.9  & 127 & 18.2 & 0.143\\
8  & 5.0 & 99  & 31.4 & 0.318 & & 8.3  & 122 & 18.1 & 0.148\\
9  & 5.4 & 92  & 28.1 & 0.305 & & 9.7  & 133 & 16.9 & 0.127\\
10 & 5.5 & 93  & 26.7 & 0.287 & & 10.3 & 144 & 16.3 & 0.113\\
11 & 6.3 & 100 & 24.5 & 0.245 & & 11.6 & 152 & 15.3 & 0.101\\
12 & 6.2 & 85  & 18.9 & 0.223 & & 11.7 & 140 & 12.5 & 0.090\\
13 & 6.8 & 91  & 17.0 & 0.187 & & 13.0 & 147 & 11.3 & 0.077\\
14 & 7.0 & 87  & 14.2 & 0.163 & & 13.2 & 143 & 10.0 & 0.070\\
15 & 7.6 & 93  & 13.2 & 0.142 & & 14.4 & 158 & 9.5  & 0.060\\
\hline
\end{tabular}

\end{center}
\end{scriptsize}
\renewcommand{\arraystretch}{1}
\end{table}

In terms of absolute numbers, a maximum throughput of up to~$E_{10} = 0.41 \ \frac{\mathrm{MDoF}}{\mathrm{s} \cdot \mathrm{core}}$ or, equivalently, a minimum solve time of~$t_{10} = 2.4 \ \frac{\mu \mathrm{s} \cdot \mathrm{core}}{\mathrm{DoF}}$  at degree~$p=3$ is achieved for the Cartesian mesh. The performance is reduced for the curvilinear mesh, with a maximum throughput of~$E_{10} = 0.22 \ \frac{\mathrm{MDoF}}{\mathrm{s} \cdot \mathrm{core}}$ and a minimum solve time of~$t_{10} = 4.6 \ \frac{\mu \mathrm{s} \cdot \mathrm{core}}{\mathrm{DoF}}$ at degree~$p=3$. The reduced performance for the curvilinear mesh compared to the Cartesian mesh can be explained by an increase in iteration counts on the one hand, and a reduced throughput of the matrix-free operator evaluation on the other hand, as summarized in Table~\ref{tab:Throughput_3D_Cartesian_vs_Curvilinear}. In addition to previous results, Table~\ref{tab:Throughput_3D_Cartesian_vs_Curvilinear} lists the effective number of fine-level matrix-vector products~$n_{10, \mathrm{mat-vec}}$. For the Cartesian mesh,~$n_{10, \mathrm{mat-vec}}\approx 100$ is obtained, i.e., solving the linear system of equations to a relative tolerance of~$\varepsilon_{10} = 10^{-10}$ corresponds to the costs of~$100$ fine-level matrix-vector products. For the curvilinear mesh,~$n_{10, \mathrm{mat-vec}}\approx 100 - 150$ is obtained with the effective number of matrix-vector products increasing for higher~$p$. To put these numbers into perspective, the cost per iteration is equivalent to $10 - 20$ fine-level matrix-vector products, while the iterative scheme performs one double-precision matrix-vector product in the CG solver, 10 single-precision matrix-vector products in the fine-level smoother in the DG space and the same number in continuous space, plus additional work on coarser levels as well as vector operations. 

\subsubsection{Comparison to state-of-the-art}

We compare the present hybrid multigrid solvers against state-of-the-art implementations from the literature in the metric~$t_{10}$: In~\cite{Stiller2017b}, the Poisson equations is solved using an interior penalty DG discretization with collocation approach on a 3D Cartesian mesh using overlapping Schwarz smoothers. A solve time of~$t_{10} \approx 7 \frac{\mu \text{s}}{\text{DoF}}$ is achieved for~$p=4$ run on a~3.1~GHz~Intel~Core~i7-5557U~CPU (one core used). Including the difference in throughput between partially loaded and fully loaded nodes according to Figure~\ref{fig:throuhgput_cube_1core_vs_1node}, the present approach can be considered significantly faster. 

In~\cite{Huismann2019}, the Poisson equation is solved on a 3D Cartesian mesh using a collocation variant of the continuous spectral element method. A solve time of~$t_{10} \approx 10 \frac{\mu \text{s}\cdot \text{core}}{\text{DoF}}$ for~$p=3$ and~$t_{10} \approx 5 \frac{\mu \text{s}\cdot \text{core}}{\text{DoF}}$ for~$p=4$ is specified in that work, where simulations have been run on a single core of a node composed of two Intel Xeon E5-2590-v3 with 12 cores each. A parallel efficiency for a fully loaded node between~$52 \%$ and $65\%$ is specified in~\cite[Table 2]{Huismann2019} for a Krylov-accelerated MG solver. This aspect needs to be taken into account and increases solve times roughly by a factor of two, see also our results in Figure~\ref{fig:throuhgput_cube_1core_vs_1node}. For moderately high polynomial degrees~$p\leq 5$, the present approach with~$t_{10} \approx 2.5 - 3 \ \frac{\mu \mathrm{s} \cdot \mathrm{core}}{\mathrm{DoF}}$ is therefore significantly more efficient, despite the fact that the implementation in~\cite{Huismann2019} uses optimizations that are restricted to Cartesian meshes and the fact that a computationally cheaper continuous finite element discretization is used. Somewhat orthogonally, the approach in~\cite{Huismann2019} is clearly faster for very large polynomial degrees such as~$p=16$, for which solve times as low as~$t_{10} \approx 1 \frac{\mu \text{s}\cdot \text{core}}{\text{DoF}}$ when using a single core are specified in that work.

In~\cite{Bastian2019}, an interior penalty DG discretization is considered for the constant coefficient Poisson problem on a 3D Cartesian geometry using block-Jacobi smoothers. A maximum performance of~$t_{8} \geq 1.33 \frac{\mu \text{s}}{\text{DoF}}$ is achieved at degree~$p=2$ on 16 cores of an Intel Xeon E5-2698v3 node, corresponding to~$t_{10} = 26.6 \frac{\mu \text{s}\cdot \text{core}}{\text{DoF}}$. Compared to the performance numbers specified above, the present approach is approximately one order of magnitude faster. It should be emphasized in this context that the smoothers used in~\cite{Bastian2019} are more complex and designed for variable-coefficient problems. At the same time, these results demonstrate that a conservative selection of smoothers with focus on robustness for potentially more complex PDEs is clearly non-optimal.

In our previous work~\cite{Kronbichler2018}, a constant-coefficient Poisson problem is solved on a 3D Cartesian geometry for an interior penalty DG discretization using a pure~$h$-multigrid approach with Chebyshev smoother of degree~$2$. A minimal solve time of~$t_{9} = 2.1 \frac{\mu \text{s} \cdot \text{core}}{\text{DoF}}$, or equivalently~$t_{10} = 2.33 \frac{\mu \text{s} \cdot \text{core}}{\text{DoF}}$, is achieved at degree~$p=4$, comparable to what is achieved in the present work, albeit on older hardware but using matrix-free kernels that are further optimized compared to the present study. In~\cite{Kronbichler2019arxiv}, an optimized code-version of this pure~$h$-multigrid method using so called cell-based face loops and merged vector operations achieves a solve time as low as~$t_{9} \geq 1.1 \frac{\mu \text{s} \cdot \text{core}}{\text{DoF}}$ (or equivalently~$t_{10} \geq 1.25 \frac{\mu \text{s} \cdot \text{core}}{\text{DoF}}$) on a hardware comparable to the present study. These optimizations have not been included in the present study since they have not been available in the~\texttt{deal.II} library by the time of writing, but indicate further performance improvements of the present hybrid multigrid methods once these optimizations are integrated.

Finally, we believe it is also very informative to compare the present DG solver with matrix-free evaluation and sum-factorization to matrix-based hybridizable DG solvers that are considered computationally efficient since the HDG approach reduces the global matrix size considerably by eliminating interior degrees of freedom and solving a linear system of equations for the trace variable living on the element boundaries only. In~\cite{Yakovlev2016}, a Helmholtz-like equation with constant coefficients is solved on a unit cube with~$9^3$ uniform hexahedral elements of degree~$p=1,\ldots,7$ and overall costs including mesh generation and setup are reported in that work using a single core on an Intel~Xeon~E7-4870 processor. A direct solver is used in that work and the authors argue that such an approach is effective in serial and for the small problem sizes considered. The wall times reported in~\cite{Yakovlev2016} range from~$5.0 \ \text{s}$ for~$p=1$, $170 \ \text{s}$ for~$p=3$, to~$16.2 \cdot 10^3 \ \text{s}$ for~$p=7$. Here, we solve the constant coefficient Poisson equation on the same mesh, which is at least as difficult to solve as a Helmholtz-like equation when using iterative solvers. We obtain wall times of~$0.59\ \text{s}$ for~$p=1$, $0.86\ \text{s}$ for~$p=3$, and~$4.5\ \text{s}$ for~$p=7$ for the whole application (including setup and postprocessing) when running the code on a single core, achieving a speed-up by a factor of~$8.5$ for~$p=1$,~$200$ for~$p=3$, and~$3600$ for~$p=7$ over the HDG results shown in~\cite{Yakovlev2016}.
Put differently, the present high-order DG results for~$p=7$ are faster than the lowest-order HDG results for~$p=1$ (on the same mesh for the same number of elements). 
These results point in a similar direction as our previous work~\cite{Kronbichler2018} where a more thorough comparative study of matrix-free DG versus matrix-based HDG methods is provided.

\subsubsection{Parameter study: influence of number of smoothing steps on iteration counts and throughput}
In this subsection, we briefly justify the choice of~$n_{\mathrm{s}}=5$ smoothing steps used for the Chebyshev smoother in previous experiments. Table~\ref{tab:chebyshev_smoothing_steps} shows the number of iterations as well as the throughput as a function of the number of smoothing steps~$n_{\mathrm{s}}$ for the Cartesian test case with a fixed polynomial degree of~$p=3$, which achieves the highest throughput in Figure~\ref{fig:sine_iterations_vs_throughput}. While the number of iterations decreases continuously for an increasing number of smoothing steps, the achieved throughput of the solver appears to only weakly depend on the number of smoothing steps, with the highest throughput achieved for a moderate number of smoothing steps. In our experience, the sweet spot is typically in the range~$n_{\mathrm{s}}=4-6$. The number of smoothing steps preferred here is higher than typically used in the literature~\cite{Sundar2015} which is due to the balance implied by mixed-precision multigrid. Overall, the results in Table~\ref{tab:chebyshev_smoothing_steps} demonstrate that there is little to gain from optimizations of the number of smoothing steps for the Chebyshev smoother.

\begin{table}[t]
\caption{Influence of number of smoothing steps~$n_{\mathrm{s}}$ on iterations~$n_{10}$ and throughput~$E_{10}$ for the~$cph$-multigrid coarsening strategy with~$p_{l-1} = \lfloor p_l/2 \rfloor$. The cube test case with Cartesian mesh is considered on a mesh with~$80^3$ elements and polynomial degree of~$p=3$~(problem size~$33\mathrm{MDoF}$).}
\label{tab:chebyshev_smoothing_steps}
\renewcommand{\arraystretch}{1.1}
\begin{scriptsize}

\begin{center}
\begin{tabular}{lrrrrrrr}
\cline{1-8}
 & \multicolumn{7}{c}{Number of smoothing steps~$n_{\mathrm{s}}$}\\
\cline{2-8}
 & 2 & 3 & 4 & 5 & 6 & 8 & 10 \\
\hline
$n_{10}$ & 11.8 & 8.4  & 6.4  & 5.3  & 5.0  & 4.3 & 3.6\\
$E_{10} \left[\frac{\mathrm{MDoF}}{\mathrm{s} \cdot \mathrm{core}}\right]$ & 0.335 & 0.367 & 0.392 & 0.404 & 0.373 & 0.368 & 0.355\\
\hline
\end{tabular}

\end{center}
\end{scriptsize}
\renewcommand{\arraystretch}{1}
\end{table}

\begin{table}[!ht]
\caption{Robustness and performance of hybrid multigrid solver for~$cph$-multigrid method with~$p_{l-1} = \lfloor p_l/2 \rfloor $ coarsening for nozzle test case. The smoother used for all experiments is Chebyshev(5,5) and the coarse-grid problem is solved iteratively to a relative tolerance of~$10^{-3}$ by the conjugate gradient method with AMG V-cycle as preconditioner.}
\label{tab:Nozzle}
\renewcommand{\arraystretch}{1.1}
\begin{scriptsize}
\begin{center}

\subtable[Iteration count~$n_{10}$]{
\begin{tabular}{lrrrrrrrrrrrrrrr}
\cline{1-16}
 & \multicolumn{15}{c}{Polynomial degree~$p$}\\
\cline{2-16}
$h$ & 1 & 2 & 3 & 4 & 5 & 6 & 7 & 8 & 9 & 10 & 11 & 12 & 13 & 14 & 15\\
\hline
$h_0$   & 5.5 & 8.4 & 11.4 & 8.6 & 10.7 & 11.8 & 12.2 & 10.5 & 11.7 & 11.8 & 12.7 & 12.0 & 12.9 & 13.7 & 14.5\\
$h_0/2$ & 8.3 & 8.8 & 12.3 & 9.6 & 11.8 & 12.5 & 13.4 & 11.5 & 12.7 & 12.0 & 12.9 & 12.8 & 13.7 & 13.8 & 14.3\\
$h_0/4$ & 8.8 & 9.8 & 13.5 & 9.8 & 11.7 & 13.5 & 14.2 & 11.7 & 12.6 & 12.5 & 12.9 & 13.7 & 13.8 & 14.6 & 14.8\\
\hline
\end{tabular}}


\subtable[Throughput~$E_{10}$ in~$\frac{\mathrm{kDoF}}{\mathrm{s} \cdot \mathrm{core}}$]{
\begin{tabular}{lrrrrrrrrrrrrrrr}
\cline{1-16}
 & \multicolumn{15}{c}{Polynomial degree~$p$}\\
\cline{2-16}
$h$ & 1 & 2 & 3 & 4 & 5 & 6 & 7 & 8 & 9 & 10 & 11 & 12 & 13 & 14 & 15\\
\hline
$h_0$   & 276  & 338  & 294 & 357 & 306 & 265 & 221  & 251 & 227  & 225  & 203  & 210  & 190  & 177  & 140 \\
$h_0/2$ & 3.30 & 36.1 & 115 & 166 & 129 & 136 & 110  & 114 & 95.1 & 90.9 & 78.8 & 73.6 & 66.6 & 62.2 & 56.4\\
$h_0/4$ & 9.28 & 82.7 & 126 & 159 & 142 & 112 & 92.8 & 106 & 94.2 & 89.4 & 84.4 & 74.9 & 71.0 & 63.8 & 59.7\\
\hline
\end{tabular}}

\subtable[Relative share of AMG coarse-grid solver in \% of wall time (`-' means costs of less than~$0.1 \%$)]{
\begin{tabular}{lrrrrrrrrrrrrrrr}
\cline{1-16}
 & \multicolumn{15}{c}{Polynomial degree~$p$}\\
\cline{2-16}
$h$ & 1 & 2 & 3 & 4 & 5 & 6 & 7 & 8 & 9 & 10 & 11 & 12 & 13 & 14 & 15\\
\hline
$h_0$   & 13.3 & 4.1 &  2.0 & 1.0 & 0.7 & 0.4 & 0.3 & 0.2 & 0.2 & 0.1 & -   & -   & -   & -   & -  \\
$h_0/2$ &  4.1 & 6.7 & 11.6 & 6.8 & 3.7 & 2.6 & 1.5 & 0.9 & 0.6 & 0.4 & 0.3 & 0.2 & 0.2 & 0.1 & 0.1\\
$h_0/4$ & 11.4 & 6.1 &  5.1 & 2.3 & 1.3 & 0.8 & 0.5 & 0.3 & 0.2 & 0.1 & 0.1 & -   & -   & -   & -  \\
\hline
\end{tabular}}

\subtable[Speed-up of~$cph$-coarsening over~$phc$-coarsening (for standard penalty factor of~$1$)]{
\begin{tabular}{lrrrrrrrrrrrrrrr}
\cline{1-16}
 & \multicolumn{15}{c}{Polynomial degree~$p$}\\
\cline{2-16}
$h$ & 1 & 2 & 3 & 4 & 5 & 6 & 7 & 8 & 9 & 10 & 11 & 12 & 13 & 14 & 15\\
\hline
$h_0$   & 0.97 & 2.05 & 1.48 & 1.70 & 1.64 & 1.47 & 1.49 & 1.57 & 1.53 & 1.60 & 1.43 & 1.47 & 1.44 & 1.35 & 1.42\\
$h_0/2$ & 1.33 & 1.42 & 2.02 & 2.04 & 1.87 & 2.04 & 1.66 & 1.71 & 1.71 & 1.62 & 1.55 & 1.57 & 1.55 & 1.49 & 1.50\\
$h_0/4$ & 1.24 & 2.19 & 1.53 & 1.87 & 1.71 & 1.48 & 1.41 & 1.62 & 1.55 & 1.54 & 1.56 & 1.47 & 1.52 & 1.46 & 1.47\\
\hline
\end{tabular}}

\subtable[Robustness of~$n_{10}$ w.r.t.~interior penalty factor~$\tau$ for mesh~$h_0/2$]{
\begin{tabular}{lrrrrrrrrrrrrrrr}
\cline{1-16}
 & \multicolumn{15}{c}{Polynomial degree~$p$}\\
\cline{2-16}
IP factor & 1 & 2 & 3 & 4 & 5 & 6 & 7 & 8 & 9 & 10 & 11 & 12 & 13 & 14 & 15\\
\hline
$10^0 \cdot \tau$ & 8.3 & 8.8 & 12.3 & 9.6 & 11.8 & 12.5 & 13.4 & 11.5 & 12.7 & 12.0 & 12.9 & 12.8 & 13.7 & 13.8 & 14.3\\
$10^1 \cdot \tau$ & 9.0 & 8.8 & 12.6 & 9.8 & 11.9 & 11.8 & 12.8 & 10.7 & 11.7 & 11.5 & 11.9 & 11.9 & 12.5 & 12.8 & 13.0\\
$10^2 \cdot \tau$ & 8.7 & 8.6 & 11.8 & 9.4 & 11.7 & 11.5 & 12.5 & 10.0 & 10.8 & 10.6 & 10.8 & 10.8 & 11.4 & 11.8 & 12.0\\
$10^3 \cdot \tau$ & 7.7 & 8.4 & 11.0 & 8.6 & 10.6 & 10.8 & 11.5 & 9.5  & 10.0 & 9.0  & 10.0 & 10.0 & 10.0 & 11.5 & 10.9\\
\hline
\end{tabular}}


\subtable[Robustness of~$n_{10}$ w.r.t.~interior penalty factor~$\tau$ for mesh~$h_0/2$ and~$phc$-coarsening strategy]{
\begin{tabular}{lrrrrrrrrrrrrrrr}
\cline{1-16}
 & \multicolumn{15}{c}{Polynomial degree~$p$}\\
\cline{2-16}
IP factor & 1 & 2 & 3 & 4 & 5 & 6 & 7 & 8 & 9 & 10 & 11 & 12 & 13 & 14 & 15\\
\hline
$10^0 \cdot \tau$ & 25.7 & 23.4 & 26.0 & 21.9 & 25.3 & 25.0 & 27.6 & 24.9 & 27.5 & 26.5 & 28.0 & 28.0 & 29.7 & 28.6 & 30.5\\
$10^1 \cdot \tau$ & 67.6 & 63.3 & 69.9 & 58.9 & 62.9 & 63.7 & 66.5 & 60.6 & 64.8 & 62.3 & 67.5 & 66.4 & 69.0 & 67.8 & 69.8\\
$10^2 \cdot \tau$ & 143  & 144  & 161  & 140  & 152  & 168  & 175  & 156  & 159  & 162  & 166  & 173  & 175  & 178  & 183 \\
$10^3 \cdot \tau$ & 187  & 208  & 228  & 201  & 231  & 262  & 291  & 261  & 293  & 301  & 330  & 339  & 372  & 373  & 387 \\
\hline
\end{tabular}}


\end{center}
\end{scriptsize}
\renewcommand{\arraystretch}{1}
\end{table}

\subsection{Nozzle}

To mimic the incompressible flow case for the nozzle problem, we prescribe a Dirichlet boundary condition with a constant value of~$1$ at the inflow boundary on the left, and a constant value of~$0$ at the outflow boundary on the right. On the walls of the nozzle geometry, homogeneous Neumann boundary conditions are prescribed. To generate a coarse grid, the nozzle domain is meshed with a minimum number of elements. The coarse grid shown in Figure~\ref{fig:geometries_and_meshes} consists of~$440$ elements and we refer to~\cite{Fehn2019} for more detailed information on the mesh generation. The coarse grid is identified as the~$h_0$ mesh in the following, and we consider meshes that are refined once ($h_0/2$) and twice ($h_0/4$) via uniform mesh refinements of the coarse mesh. A cubic mapping is used for all computations for a high-order representation of the geometry which is described via manifold descriptions. For polynomial degrees from~$p=1,\ldots,15$, the problem size ranges from~$3.5 \cdot 10^3 - 1.8 \cdot 10^6$ unknowns for mesh~$h_0$,~$2.8 \cdot 10^4 - 1.4 \cdot 10^7$ unknowns for~$h_0/2$, and~$2.3 \cdot 10^5 - 1.2 \cdot 10^8$ unknowns for~$h_0/4$. Computations on mesh~$h_0$ are performed on one core due to the small problem size, on mesh~$h_0/2$ on one node~(48 cores), and on mesh~$h_0/4$ on two nodes~(96 cores).

Table~\ref{tab:Nozzle} summarizes the numerical results for the nozzle geometry of the FDA benchmark where we focus on the~$cph$-multigrid method. In terms of iteration counts, mesh independent convergence is observed, and a slight increase in the number of iterations for large~$p$ in agreement with previous results. Compared to the curvilinear mesh for the cube problem, the number of iterations is larger, explaining the reduced throughput~$E_{10}$ compared to the results on the curvilinear mesh for the cube geometry in Table~\ref{tab:Throughput_3D_Cartesian_vs_Curvilinear}. An increased throughput is measured on the coarse mesh~$h_0$, since the computations are performed on a single core, see also Figure~\ref{fig:throuhgput_cube_1core_vs_1node}. On the finer meshes, the throughput is small for low polynomial degrees. This is due to the fact the problem size covers a broad range from a very low to high workload per core when going from~$p=1$ to~$p=15$ for a fixed number of elements (in contrast, the number of elements has been adapted for the cube test case to obtain a similar problem size for all~$p$).

Table~\ref{tab:Nozzle} also lists the relative share of the AMG coarse-grid solver in \% of the overall wall time required by the linear solver. The coarse-grid solver accounts for up to~$13 \%$ of the computational costs for linear shape functions~($p=1$), and becomes negligible in terms of computational costs for increasing polynomial degree and finer meshes. By the use of hybrid multigrid methods, the overall computational efficiency of the method is determined by the fast matrix-free operator evaluation on the finest levels as intended.

The computationally efficient~$cph$- and~$chp$-coarsening strategies show a similar performance for the nozzle test case both in terms of iteration counts and computational costs, so that no significant advantage of one over the other method could be identified. As shown in Table~\ref{tab:Nozzle}, the~$cph$-multigrid method is more efficient than the~$phc$-method for all polynomial degrees and meshes considered for the nozzle problem. Robustness w.r.t.~the interior penalty factor is obtained for the~$cph$-multigrid method (and similarly for~$chp$-coarsening), while a strong increase in iteration counts is observed in case of~$phc$-coarsening.

\subsection{Lung}

A specialized mesh generator has been developed to be able to mesh complex lung geometries with purely hexahedral elements, see Figure~\ref{fig:geometries_and_meshes}. The patient-specific geometry of the first three generations is obtained from a segmentation of MRI scans, while higher airway generations are constructed using a recursive tree growing algorithm that mimics the true anatomy of the preterm infant and respects anatomical length and diameter ratios of airways reported for the preterm infant~\cite{Roth2018}. In a first step, a 3D cylinder tree is created, which is subsequently deformed according to the patient-specific geometry of upper airway generations obtained from MR images and described via B-splines. When refining the mesh, new nodes are placed correctly on the patient-specific geometry. A tri-linear mapping of the geometry is used in the present study. Also for the lung test case, the application in mind is the solution of the pressure Poisson equation as part of an incompressible Navier--Stokes solver. Therefore, we prescribe a Dirichlet boundary value of~$1$ at the upper boundary and homogeneous Dirichlet boundary conditions at all outlets where the airways that are resolved by this lung model end. Homogeneous Neumann boundary conditions are prescribed on all airway walls.

\begin{table}[!ht]
\caption{Robustness and performance of hybrid multigrid solver for~$cph$-multigrid method with~$p_{l-1} = \lfloor p_l/2 \rfloor $ coarsening for lung test case. The smoother used for all experiments is Chebyshev(5,5) and the coarse-grid problem is solved iteratively to a relative tolerance of~$10^{-1}$ by the conjugate gradient method with AMG V-cycle with Chebyshev(3,3) smoother as preconditioner.}
\label{tab:Lung}
\renewcommand{\arraystretch}{1.1}
\begin{scriptsize}
\begin{center}

\subtable[Iteration count~$n_{10}$]{
\begin{tabular}{lrrrrrrrrrrrrrrr}
\cline{1-16}
 & \multicolumn{15}{c}{Polynomial degree~$p$}\\
\cline{2-16}
$h$ & 1 & 2 & 3 & 4 & 5 & 6 & 7 & 8 & 9 & 10 & 11 & 12 & 13 & 14 & 15\\
\hline
$h_0$   & 12.3 & 19.3 & 27.6 & 19.9 & 26.8 & 29.4 & 30.7 & 27.9 & 32.6 & 32.9 & 36.5 & 36.4 & 39.5 & 39.4 & 41.9\\
$h_0/2$ & 17.6 & 20.0 & 28.9 & 22.6 & 29.6 & 30.8 & 36.8 & 33.7 & 38.6 & 38.0 & 42.5 & 40.7 & 43.0 & 43.9 & 45.7\\
$h_0/4$ & 18.5 & 21.0 & 30.9 & 25.9 & 32.6 & 34.6 & 39.7 & 35.9 & 40.7 & 40.5 & 44.4 & 43.7 & 47.9 & 47.4 & 51.5\\
\hline
\end{tabular}}

\subtable[Throughput~$E_{10}$ in~$\frac{\mathrm{kDoF}}{\mathrm{s} \cdot \mathrm{core}}$]{
\begin{tabular}{lrrrrrrrrrrrrrrr}
\cline{1-16}
 & \multicolumn{15}{c}{Polynomial degree~$p$}\\
\cline{2-16}
$h$ & 1 & 2 & 3 & 4 & 5 & 6 & 7 & 8 & 9 & 10 & 11 & 12 & 13 & 14 & 15\\
\hline
$h_0$   & 5.47 & 36.8 & 53.2 & 80.9 & 64.7 & 57.3 & 49.0 & 49.8 & 40.9 & 37.8 & 33.2 & 31.3 & 27.9 & 26.5 & 23.5\\
$h_0/2$ & 39.8 & 90.4 & 68.3 & 76.3 & 57.4 & 52.1 & 42.0 & 44.2 & 38.0 & 37.2 & 32.3 & 31.2 & 28.2 & 25.3 & 22.6\\
$h_0/4$ & 24.2 & 77.8 & 58.1 & 61.8 & 48.2 & 43.6 & 37.3 & 40.1 & 35.6 & 34.2 & 30.6 & 29.3 & 25.2 & 23.5 & 20.4\\
\hline
\end{tabular}}

\subtable[Relative share of AMG coarse-grid solver in \% of wall time (`-' means costs of less than~$0.1 \%$)]{
\begin{tabular}{lrrrrrrrrrrrrrrr}
\cline{1-16}
 & \multicolumn{15}{c}{Polynomial degree~$p$}\\
\cline{2-16}
$h$ & 1 & 2 & 3 & 4 & 5 & 6 & 7 & 8 & 9 & 10 & 11 & 12 & 13 & 14 & 15\\
\hline
$h_0$   & 15.0 & 21.7 & 18.5 & 8.5 & 5.6 & 3.4 & 2.1 & 1.4 & 0.9 & 0.7 & 0.5 & 0.4 & 0.3 & 0.2 & 0.2\\
$h_0/2$ & 19.7 &  6.3 &  2.7 & 1.2 & 0.8 & 0.4 & 0.3 & 0.2 & 0.1 &   - &   - &   - &   - &   - &   -\\
$h_0/4$ & 18.1 &  9.8 &  3.9 & 1.9 & 1.1 & 0.7 & 0.4 & 0.3 & 0.2 & 0.1 & 0.1 &   - &   - &   - &   -\\
\hline
\end{tabular}}

\subtable[Speed-up of~$cph$-coarsening over~$phc$-coarsening (for standard penalty factor of~$1$)]{
\begin{tabular}{lrrrrrrrrrrrrrrr}
\cline{1-16}
 & \multicolumn{15}{c}{Polynomial degree~$p$}\\
\cline{2-16}
$h$ & 1 & 2 & 3 & 4 & 5 & 6 & 7 & 8 & 9 & 10 & 11 & 12 & 13 & 14 & 15\\
\hline
$h_0$   & 0.72 & 1.80 & 1.51 & 1.88 & 1.62 & 1.36 & 1.47 & 1.59 & 1.55 & 1.47 & 1.51 & 1.47 & 1.48 & 1.47 & 1.51\\
$h_0/2$ & 1.83 & 2.18 & 1.51 & 1.60 & 1.34 & 1.32 & 1.27 & 1.35 & 1.35 & 1.35 & 1.36 & 1.37 & 1.48 & 1.35 & 1.39\\
$h_0/4$ & 1.77 & 2.22 & 1.52 & 1.67 & 1.37 & 1.27 & 1.29 & 1.40 & 1.41 & 1.30 & 1.34 & 1.36 & 1.38 & 1.38 & 1.38\\
\hline
\end{tabular}}

\subtable[Robustness of~$n_{10}$ w.r.t.~interior penalty factor~$\tau$ for mesh~$h_0/2$]{
\begin{tabular}{lrrrrrrrrrrrrrrr}
\cline{1-16}
 & \multicolumn{15}{c}{Polynomial degree~$p$}\\
\cline{2-16}
IP factor & 1 & 2 & 3 & 4 & 5 & 6 & 7 & 8 & 9 & 10 & 11 & 12 & 13 & 14 & 15\\
\hline
$10^0 \cdot \tau$ & 17.6 & 20.0 & 28.9 & 22.6 & 29.6 & 30.8 & 36.8 & 33.7 & 38.6 & 38.0 & 42.5 & 40.7 & 43.0 & 43.9 & 45.7\\
$10^1 \cdot \tau$ & 16.9 & 19.6 & 28.5 & 20.9 & 27.8 & 29.4 & 32.6 & 32.6 & 34.6 & 33.9 & 36.9 & 35.8 & 39.4 & 38.9 & 40.0\\
$10^2 \cdot \tau$ & 16.4 & 18.6 & 27.2 & 19.9 & 25.7 & 26.8 & 29.6 & 28.3 & 32.7 & 31.8 & 34.4 & 33.7 & 34.9 & 34.0 & 37.0\\
$10^3 \cdot \tau$ & 16.3 & 17.9 & 24.7 & 18.8 & 25.0 & 25.8 & 27.0 & 27.0 & 28.9 & 28.9 & 29.3 & 28.6 & 31.4 & 32.6 & 34.3\\
\hline
\end{tabular}}


\subtable[Robustness of~$n_{10}$ w.r.t.~interior penalty factor~$\tau$ for mesh~$h_0/2$ and~$phc$-coarsening strategy]{
\begin{tabular}{lrrrrrrrrrrrrrrr}
\cline{1-16}
 & \multicolumn{15}{c}{Polynomial degree~$p$}\\
\cline{2-16}
IP factor & 1 & 2 & 3 & 4 & 5 & 6 & 7 & 8 & 9 & 10 & 11 & 12 & 13 & 14 & 15\\
\hline
$10^0 \cdot \tau$ & 52.1 & 47.9 & 54.7 & 46.0 & 54.7 & 56.7 & 66.9 & 63.8 & 75.7 & 73.3 & 83.5 & 78.6 & 87.6 & 81.5 & 88.5\\
$10^1 \cdot \tau$ & 234  & 178  & 200  & 152  & 143  & 139  & 169  & 155  & 176  & 165  & 184  & 174  & 192  & 175  & 191\\
$10^2 \cdot \tau$ & 859  & 604  & 1158 & 1139 & 671  & 587  & 514  & 708  & 435  & 409  & 458  & 429  & 471  & 466  & 511\\
$10^3 \cdot \tau$ & 1579 & 1535 & 1589 & 1547 & 4169 & 1078 & 1069 & 1803 & 3729 & 1304 & 1177 & 1116 & 1242 & 1248 & 1365\\
\hline
\end{tabular}}


\end{center}
\end{scriptsize}
\renewcommand{\arraystretch}{1}
\end{table}

As for the nozzle problem, we solve the problem on the coarse mesh labeled~$h_0$, and consider two finer meshes~$h_0/2$ and~$h_0/4$ obtained via uniform mesh refinement. In the following, we show results for the mesh resolving 8 generations of the lung with a coarse mesh consisting of~$9396$ elements, see Figure~\ref{fig:geometries_and_meshes}. The lung mesh contains bad-aspect-ratio elements so that this test case represents more practical, difficult problems. For polynomial degrees from~$p=1,\ldots,15$, the problem size ranges from~$7.5 \cdot 10^4 - 3.8 \cdot 10^7$ unknowns for mesh~$h_0$,~$6.0 \cdot 10^5 - 3.1 \cdot 10^8$ unknowns for~$h_0/2$, and~$4.8 \cdot 10^6 - 2.5 \cdot 10^9$ unknowns for~$h_0/4$. Computations on meshes~$h_0$ and~$h_0/2$ are done on one fat compute node~(48 cores) and computations on mesh~$h_0/4$ on 8 fat nodes~(384 cores). For the lung test case, we observed that the AMG coarse grid preconditioner with ILU smoother lacks robustness with respect to the number of cores. Hence, we use a Chebyshev(3,3) smoother for the AMG coarse grid preconditioner.

Table~\ref{tab:Lung} lists the results for the lung test case mainly focusing on the~$cph$-multigrid method. The number of iterations~$n_{10}$ increase slightly on finer meshes, and more strongly for increasing~$p$. The number of iterations is highest for the lung test case explaining the reduction in throughput~$E_{10}$ compared to the results for the cube geometry with curvilinear mesh. The~$cph$-multigrid method is faster than the~$phc$-multigrid method for all polynomial degrees due to a significant reduction in iteration counts. Regarding the interior penalty parameter, robustness is obtained for~$cph$- and~$chp$-coarsening, and a strong increase in iterations counts is observed, e.g., in case of~$phc$- and~$hpc$-coarsening. The~$cph$- and~$chp$-methods (and similarly the~$phc$- and~$hpc$-methods) perform similarly for the lung problem with a small advantage for~$ph$-type approaches due to slightly smaller iteration counts in agreement with the results in Figure~\ref{fig:sine_iterations_vs_throughput} for the cube problem. The costs of the AMG coarse-grid solver are negligible for higher order methods and the coarse-grid solver does also not form a bottleneck for the lowest polynomial degrees, demonstrating a proper design of the present multigrid algorithms by the use of hybrid coarsening strategies.

\begin{table}[!ht]
\caption{Performance of pure~$p$-\textbf{multigrid} method with~$p_{l-1} = \lfloor p_l/2 \rfloor $ coarsening for lung test case versus hybrid~$phc$-multigrid method. The smoother used for all~$p$-MG experiments is Chebyshev(5,5) and the coarse-grid problem is solved iteratively to a relative tolerance of~$10^{-1}$ by the conjugate gradient method with AMG V-cycle with Chebyshev(3,3) smoother as preconditioner.}
\label{tab:Lung_HybridMultigrid_Versus_StandardPMultigrid}
\renewcommand{\arraystretch}{1.1}
\begin{scriptsize}
\begin{center}

\subtable[Relative share of AMG coarse-grid solver in \% of wall time for pure~$p$-multigrid approach]{
\begin{tabular}{lrrrrrrrrrrrrrrr}
\cline{1-16}
 & \multicolumn{15}{c}{Polynomial degree~$p$}\\
\cline{2-16}
$h$ & 1 & 2 & 3 & 4 & 5 & 6 & 7 & 8 & 9 & 10 & 11 & 12 & 13 & 14 & 15\\
\hline
$h_0$   & 76.5 & 70.6 & 65.7 & 52.1 & 39.4 & 30.4 & 18.1 & 11.5 & 7.8  & 5.5  & 3.7 & 3.8 & 2.9 & 1.8 & 1.5\\
$h_0/2$ & 98.4 & 91.9 & 78.9 & 69.9 & 53.3 & 35.0 & 17.8 & 18.9 & 15.8 & 10.7 & 8.0 & 5.2 & 3.3 & 3.1 & 3.1\\
$h_0/4$ & 97.6 & 92.5 & 82.2 & 73.6 & 62.7 & 43.5 & 20.1 & 19.1 & 17.9 & 13.3 & 8.9 & 6.5 & 5.0 & 5.7 & 4.2\\
\hline
\end{tabular}}

\subtable[Speed-up of~$phc$-multigrid over pure~$p$-multigrid]{
\begin{tabular}{lrrrrrrrrrrrrrrr}
\cline{1-16}
 & \multicolumn{15}{c}{Polynomial degree~$p$}\\
\cline{2-16}
$h$ & 1 & 2 & 3 & 4 & 5 & 6 & 7 & 8 & 9 & 10 & 11 & 12 & 13 & 14 & 15\\
\hline
$h_0$   & 1.35 & 3.01 & 2.20 & 1.75 & 1.46 & 1.35 & 1.24 & 1.09 & 1.06 & 1.04 & 1.02 & 1.02 & 1.03 & 0.98 & 1.00\\
$h_0/2$ & 6.59 & 8.50 & 3.86 & 2.88 & 2.03 & 1.46 & 1.20 & 1.22 & 1.18 & 1.12 & 1.07 & 1.06 & 1.03 & 1.03 & 1.04\\
$h_0/4$ & 5.33 & 8.60 & 5.12 & 3.02 & 2.50 & 1.66 & 1.22 & 1.22 & 1.22 & 1.15 & 1.10 & 1.06 & 1.05 & 1.06 & 1.04\\
\hline
\end{tabular}}

\end{center}
\end{scriptsize}
\renewcommand{\arraystretch}{1}
\end{table}

\begin{table}[!ht]
\caption{Performance of pure~$h$-\textbf{multigrid} method for lung test case versus hybrid~$hpc$-multigrid method. FGMRES(30) is used as outer Krylov solver preconditioned by an~$h$-MG V-cycle with Chebyshev(5,5) smoother and the coarse-grid problem is solved iteratively to a relative tolerance of~$10^{-1}$ by the conjugate gradient method with point-Jacobi as preconditioner.}
\label{tab:Lung_HybridMultigrid_Versus_StandardHMultigrid}
\renewcommand{\arraystretch}{1.1}
\begin{scriptsize}
\begin{center}

\subtable[Relative share of coarse-grid solver in \% of wall time for pure~$h$-multigrid approach]{
\begin{tabular}{lrrrrrrrrrrrrrrr}
\cline{1-16}
 & \multicolumn{15}{c}{Polynomial degree~$p$}\\
\cline{2-16}
$h$ & 1 & 2 & 3 & 4 & 5 & 6 & 7 & 8 & 9 & 10 & 11 & 12 & 13 & 14 & 15\\
\hline
$h_0$   & 85.4 & 98.5 & 99.7 & 99.5 & 100  & 99.8 & 99.9 & 100  & 99.6 & 100  & 99.9 & 100  & 100  & 100  & 100 \\
$h_0/2$ & 64.3 & 68.7 & 73.0 & 74.9 & 76.4 & 86.0 & 88.5 & 88.3 & 90.2 & 91.2 & 94.1 & 94.4 & 93.5 & 94.6 & 95.4\\
$h_0/4$ & 55.8 & 53.8 & 44.9 & 42.1 & 40.7 & 46.5 & 53.0 & 56.2 & 60.7 & 69.4 & 67.2 & 71.4 & 72.3 & 69.5 & 74.7\\
\hline
\end{tabular}}

\subtable[Speed-up of~$hpc$-multigrid over pure~$h$-multigrid]{
\begin{tabular}{lrrrrrrrrrrrrrrr}
\cline{1-16}
 & \multicolumn{15}{c}{Polynomial degree~$p$}\\
\cline{2-16}
$h$ & 1 & 2 & 3 & 4 & 5 & 6 & 7 & 8 & 9 & 10 & 11 & 12 & 13 & 14 & 15\\
\hline
$h_0$   & 3.26 & 4.84 & 9.78 & 13.8 & 19.7 & 24.9 & 28.8 & 35.4 & 36.7 & 44.5 & 46.7 & 50.7 & 52.4 & 56.3 & 58.4\\
$h_0/2$ & 2.15 & 3.09 & 3.93 & 4.15 & 5.24 & 9.25 & 11.1 & 10.3 & 12.2 & 13.9 & 21.3 & 21.0 & 18.4 & 21.6 & 25.2\\
$h_0/4$ & 1.55 & 2.04 & 2.34 & 2.38 & 2.37 & 2.74 & 3.10 & 3.35 & 3.81 & 4.90 & 4.40 & 4.99 & 5.58 & 5.00 & 6.12\\
\hline
\end{tabular}}

\end{center}
\end{scriptsize}
\renewcommand{\arraystretch}{1}
\end{table}

As shown in Tables~\ref{tab:Lung_HybridMultigrid_Versus_StandardPMultigrid} and~\ref{tab:Lung_HybridMultigrid_Versus_StandardHMultigrid}, this is not the case for pure~$p$-multigrid and pure~$h$-multigrid methods frequently used in the literature. For the~$h$-multigrid approach, the AMG coarse grid solver is very ineffective when applied to high-order discretizations and the memory of the fat compute nodes soon becomes exhausted when going to higher order, highlighting severe limitations of matrix-based approaches. More simple coarse grid solvers such as the Chebyshev iteration or a conjugate gradient iteration with point-Jacobi preconditioning are much more efficient. The latter coarse grid solver (CG with point-Jacobi) was identified as the most efficient coarse grid solver and is used in Table~\ref{tab:Lung_HybridMultigrid_Versus_StandardHMultigrid}. Since the convergence behavior of this coarse grid solver is rather slow in terms of iteration counts, it is essential to use FGMRES as outer Krylov solver in this case.

Regarding the~$p$-multigrid results in Table~\ref{tab:Lung_HybridMultigrid_Versus_StandardPMultigrid}, the AMG coarse-grid solver constitutes the main bottleneck for moderately large polynomial degrees~$p \leq 5$. Only for very large polynomial degrees~$p=10-15$ the pure~$p$-multigrid method allows enough coarsening to make sure that the coarse-grid solver becomes negligible in terms of costs.~\footnote{As mentioned previously, we apply the AMG coarse-grid solver in a black-box fashion without performance optimizations, e.g., by additional parameter studies. One might therefore argue that the performance of the AMG solver might have potential for further improvements. At the same time, one can argue that a hybrid multigrid method with more aggressive coarsening and negligible costs for the coarse-grid problem is advantageous as it eliminates the need to tune parameters related to AMG.} As expected theoretically, the pure~$h$-multigrid approach behaves orthogonally to the~$p$-multigrid approach. For one or two mesh refinement levels (what we believe is typical of practical applications), a significant share of the overall costs is still due to the coarse-grid solver for all polynomial degrees~$1 \leq p \leq 15$. The hybrid~$phc$- and~$hpc$-multigrid methods outperform the pure~$p$- and~$h$-multigrid methods for all polynomial degrees and meshes, where the large speed-up factors originate from the fact that the bottleneck of the coarse-grid solver is removed. Note that we use a~$p_{l-1} = \lfloor p_l/2 \rfloor$ coarsening here, so that the reported speed-up is conservative in terms of the available~$p$-coarsening types used in the literature. Compared to the~$phc$- and~$hpc$-coarsening used in Tables~\ref{tab:Lung_HybridMultigrid_Versus_StandardPMultigrid} and~\ref{tab:Lung_HybridMultigrid_Versus_StandardHMultigrid}, further speed-up by up to a factor of two can be achieved by performing the~$c$-transfer on the finest level as shown in Table~\ref{tab:Lung}.

According to these results and with reference to Table~\ref{tab:HybridMultigridApproaches}, we conclude that the different methods proposed previously in the literature are optimal only in certain regimes, and that the hybrid multigrid techniques with sum-factorized matrix-free implementation discussed here become mandatory in order to achieve a versatile PDE solver efficient for a wide range of problems and spatial resolution parameters~$h$ and~$p$.

\section{Conclusion and outlook}\label{sec:Summary}
The present work presents hybrid multigrid techniques for high-order DG discretizations, i.e., multigrid coarsening strategies that exploit all levels of geometric, polynomial, and algebraic coarsening. In addition, a transfer from discontinuous to continuous finite element spaces is performed. We have discussed the relevant design choices in the context of hybrid multigrid methods and conducted performance comparisons for various multigrid methods and different types of~$p$-coarsening in the metric of computational costs. Optimal-complexity matrix-free operator evaluation is exploited on all multigrid levels, smoothers, and transfer operators except for the coarse-grid solver. The performance is further improved by the use of mixed-precision multigrid. Our results can be summarized as follows: i) a~$p_{l-1} = \lfloor p_l/2 \rfloor $ coarsening strategy that reduces the number of unknowns roughly in factors of~$2^d$ from one level to the next performs better than other~$p$-coarsening types that reduce the polynomial degree by one until the lowest degree is reached, or directly from high-order to the lowest polynomial degree within one level. ii) Performing the~$c$-transfer from discontinuous to continuous space at the fine level is superior to an alternative~$c$-transfer performed at the coarse level before the coarse-grid solver is invoked. Moreover, this approach yields a multigrid algorithm with iteration counts independent of the penalty factor of the interior penalty method. The~$cph$- and~$chp$-multigrid methods are identified as most promising. iii) By the development of hybrid multigrid methods that exploit all possibilities of~$h$-,~$p$-, and~$c$-coarsening, the bottleneck of expensive coarse-grid solvers is significantly relaxed that would otherwise dominate overall computational costs.

We believe that the highest potential for further performance improvements lies in the development of multigrid smoothers that are robust for anisotropic problems and that can be realized in an entirely matrix-free way. In the future, we want to extend the hybrid multigrid methods proposed here towards~$hp$-adaptivity. As part of future work, we also plan an in-depth investigation of the parallel scalability of the present hybrid multigrid methods, including alternative AMG coarse-grid solvers. 

\appendix

\section*{Acknowledgments}
The research presented in this paper was partly funded by the German Research Foundation (DFG) under the project ``High-order discontinuous Galerkin for the EXA-scale'' (ExaDG) within the priority program ``Software for Exascale Computing'' (SPPEXA), grant agreement no. KR4661/2-1 and WA1521/18-1. The authors gratefully acknowledge the Gauss Centre for Supercomputing e.V.~(\texttt{www.gauss-centre.eu}) for funding this project by providing computing time on the GCS Supercomputer SuperMUC-NG at Leibniz
Supercomputing Centre (LRZ, \texttt{www.lrz.de}) through project id pr83te.


\bibliography{paper}

\end{document}